\documentclass[aps,reprint,amsmath,amssymb,showpacs,showkeys]{revtex4-1}

\usepackage{graphicx}  
\usepackage{natbib}  
\usepackage{amsmath}  
\usepackage{hyperref}  
\usepackage[top=1.5in, bottom=1.5in, left=1in, right=1in]{geometry}
\usepackage{caption}
\usepackage{amsfonts}
\usepackage{xcolor}
\usepackage{booktabs}

\newcommand{\bmt}[1]{\mbox{\boldmath $#1$}}
\newcommand{\w}{{\bf w}}
\begin{document}\sloppy  

\title{Identifying and Addressing Nonstationary LISA Noise}

\author{Matthew C. Edwards$^{1, 2}$, Patricio Maturana-Russel$^{1,3}$,
  Renate Meyer$^1$, Jonathan Gair$^{2, 4}$, Natalia Korsakova$^5$,
  Nelson Christensen$^5$} \affiliation{$^1$ Department of Statistics,
  University of Auckland, Auckland, New Zealand \\ $^2$ School of
  Mathematics, University of Edinburgh, Edinburgh, United
  Kingdom\\ $^3$ Department of Mathematical Sciences, Auckland
  University of Technology, Auckland, New Zealand \\ $^4$ Albert
  Einstein Institute, Max Planck Institute for Gravitational Physics,
  Potsdam, Germany \\ $^5$ Universit\'{e} C\^{o}te d'Azur,
  Observatoire de C\^{o}te d'Azur, CNRS, Artemis, Nice, France}

\begin{abstract}

  We anticipate noise from the Laser Interferometer Space Antenna
  (LISA) will exhibit nonstationarities throughout the duration of its
  mission due to factors such as antenna repointing,
  cyclostationarities from spacecraft motion, and glitches as
  highlighted by LISA Pathfinder.  In this paper, we use a surrogate
  data approach to test the stationarity of a time series
  which does not rely on the Gaussianity assumption.
  The main goal is to identify noise nonstationarities in the future
  LISA mission.  This will be necessary for determining how often the
  LISA noise power spectral density (PSD) will need to be updated for
  parameter estimation routines.  We conduct a thorough simulation
  study illustrating the power/size of various versions of the
  hypothesis tests, and then apply these approaches to differential
  acceleration measurements from LISA Pathfinder.  We also develop a
  data analysis strategy for addressing nonstationarities in the LISA
  PSD, where we update the noise PSD over time, while simultaneously
  conducting parameter estimation, with a focus on planned data gaps.
  
\end{abstract}


\maketitle


\section{Introduction}\label{sec:intro}

The Laser Interferometer Space Antenna (LISA) is a planned space-based
gravitational wave (GW) mission with an expected launch in 2034 led by
the European Space Agency (ESA) \citep{amaroseoane2017laser}.  The aim
of this mission is to observe GW signals in the millihertz band which
among others include astrophysical objects such as galactic white
dwarf binaries \citep{carre:2010}, massive and supermassive black hole
binaries \citep{sesana:2005}, and extreme mass ratio inspirals (EMRIs)
\citep{chua:2017}.  LISA will consist of a set of three spacecrafts
arranged into an ``equilateral'' triangle, each separated by $L = 2.5
\times 10^6~\mathrm{km}$, connected with a laser link.  The LISA
constellation will cartwheel in an Earth-trailing heliocentric orbit
around the Sun at an angle of 20 degrees between the Sun and Earth.

We expect LISA noise will be nonstationary in numerous ways.  For
example, as the spacecrafts will not always be able to point in the
same direction towards Earth for us to receive data, there will be
planned communication interruptions (or gaps), where the antennae will be
repointed to adjust the beam \citep{carre:2010, baghi:2019}.  This
means physically moving the antennae, which will create noise.
Another subtle effect of the repointing is that the distribution of
mass near the test mass will change, which might affect the gravity
gradient noise, leading to a change in acceleration noise
\citep{Purdue_2007,PhysRevD.99.122003}.  Controls may need to actively
hold the proof mass using electrostatic actuation, which may lead to
charging of the proof mass, and a change in the state of the noise
\citep{Baker:2019pnp,Pollack:2009ef,armano2019lisa}.

Cyclostationarities are also expected in LISA, for example, due to the
cartwheeling motion and orbits of the satellites.  As LISA does not
have uniform sensitivity in the sky and is more sensitive in the
direction perpendicular to the plane of the constellation, there will
be higher amplitude confusion noise when pointing to the line of sight
of the galactic centre as this is where a large amount of galactic
white dwarf binaries are located \citep{Lamberts:2019nyk}.  In
addition, LISA has a periodic orbit around the Sun, and
pseudo-periodic solar activity can lead to cyclostationary noise
\citep{Adams:2010vc,Adams:2013qma}.

LISA Pathfinder (LPF) was an ESA satellite whose goal was to
demonstrate the technology for the future LISA mission
\citep{lpf:2016}.  Glitches in differential acceleration measurements
$\Delta g$ have been analyzed in previous studies, occurring at a rate
of one glitch per two days
\citep{lpf:2016,PhysRevLett.120.061101}. As LISA will
have a similar architecture to LPF, we expect glitches as another form
of nonstationarity in the future mission \citep{Robson:2018jly}.

To understand exactly what it means to have nonstationary noise, first
we must discuss precisely what a stationary process is.  A (weakly)
stationary time series $\mathbf{Y} = (Y_1, Y_2, \ldots, Y_n)^\top$ is
a stochastic process that has constant and finite mean and variance
over time, i.e.,
\begin{eqnarray*}
\mathbb{E}[Y_t] &=& \mu < \infty,\\
\mathrm{Var}[Y_t] &=& \sigma^2 < \infty,
\end{eqnarray*}
for all $t$, and an autocovariance function $\gamma(.)$ that depends
only on the time lag $s$ \citep{brockwell:1991}.  That is, for a
zero-mean weakly stationary process, the autocovariance function has
the form
\begin{equation*}
  \label{eq:autocovariance}
  \gamma(s) = \mathbb{E}[Y_t Y_{t+s}], \quad \forall t,
\end{equation*}
where $\mathbb{E}[.]$ is the expected value operator, and $t$
represents time.  Note that the PSD function is the Fourier transform
of the autocovariance function.  

Nonstationarities in a time series can therefore come in the form of a
trend, heteroskedasticity, or time-varying autocorrelations (or PSDs).
One can also consider amplitude modulation (AM) and frequency
modulation (FM) to be forms of nonstationarity.  In this paper, we are
interested in a time-varying PSD structure, where we want to identify
and handle this type of nonstationarity.  To this end, we propose two
hypothesis tests to identify whether a time series is stationary in
terms of its PSD, which will be described in Sections~\ref{sec:vocal}
and \ref{sec:somed}.  Further, we have developed an analysis strategy
for dealing with nonstationary LISA noise, where we update the
estimate of the noise PSD over time, rather than fixing it and
assuming stationarity.  It is worth noting that in the context of
Laser Interferometer Gravitational-Wave Observatory (LIGO) data
analysis, fluctuations in the PSD can bias parameter estimates
\citep{aasi:2013,LIGOScientific:2019hgc,biscoveanu:2020}.  Here, we
are particularly interested in the gap problem \citep{carre:2010,
  baghi:2019}, where we believe satellite repointing could temporarily
change the noise structure of the LISA satellites.

Common approaches to testing the stationarity of a time series are the
so-called \textit{unit root tests}, including the Augmented
Dickey-Fuller (ADF) test \citep{adf_test:1984}, Phillips-Perron (PP)
test \citep{pp_test:1988}, and the Kwiatkowski-Phillips-Schmidt-Shin
(KPSS) test \citep{kpss_test:1992} for detecting a
  particular type of nonstationarity, namely a unit root
  autoregressive process. The behaviour of these unit root tests
  strongly depends on the long-run variance estimator used for
  rescaling the test statistic and they often fail to control the
  size, i.e.\ falsely reject stationarity too often for stationary
  time series with strong autocorrelation \citet{muller:2005}.  Unit
root tests have been noted in the GW literature by \citet{romano:2017}
to not be of particular value as GW noise generally exhibits high
autocorrelation with roots close to the unit circle.
Moreover, these tests depend on the assumption of
  Gaussianity which may not be appropriate for GW data in the presence
  of glitches.

A purely visual test to check whether the periodograms
  change over time is based on the spectrogram by dividing the time
  series into smaller segments, and visualizing the successive
  segment-based periodograms. These form the starting point for
  formal \textit{spectral analysis tests} that consider evolutionary
(or time-varying) spectral estimates using time-frequency
representations of the data. They share the common
  principle of comparing statistics based on adjacent segments. The
most notable of these are the wavelet tests of \citet{vonSachs:1998}
and \citet{nason:2000}, where the authors propose using Haar wavelets
of time-varying periodograms to test for covariance stationarity, and
the Priestley-Subba Rao test \citep{priestly:1969} which tests the
uniformity of a set of evolutionary spectra at different time
intervals, and is similar to a two-factor analysis of variance
(ANOVA). The wavelet test and Priestley-Subba Rao
  test use the asymptotic distribution of their test statistic under
  various assumptions on the local spectra which might be difficult to
  verify in any particular situation and often rely on Gaussian
  distributions, thus failing to control the size for heavy-tailed
  distributions. The Priestley-Subba Rao test requires the
  independence of time-frequency bins which may lead to stationarity
  decision errors due to biased estimations.
  In the context of GW data analysis for LIGO and Virgo,
  \citet{LIGOScientific:2019hgc} visualized potential nonstationarity
  of LIGO noise time series by a scalogram showing the amplitudes of
  wavelet basis functions at each discrete time and frequency. After
  prewhitening the data, the sum of squares of wavelet amplitudes
  would have a chi-squared distribution when applied to stationary
  Gaussian noise.  Then, an Anderson-Darling test
  \citep{anderson:1954} was applied to test against deviations from
  this chi-squared distribution.  Its performance will depend
  critically on the assumption of Gaussianity and the spectral density
  estimate used for pre-whitening.  Therefore, the development of
  stationarity tests  against the alternative of a
  time-varying PSD that do not rely on Gaussian assumptions is
  important for practical analysis of GW data.

To avoid reliance on restrictive assumptions to
  derive the asymptotic distribution of the test statistic under the
  null hypothesis, various \textit{resampling} approaches for testing
the stationarity of a time series have also been introduced.
One such approach by \citet{swanepoel:1986} uses a modification of the
bootstrap of \citet{efron:1979} to test the equality of two spectral
densities from two independent time series. This
  approach still depends on parametric assumptions as autoregressive
  models are fitted to the data in each segment and the bootstrap is
  based on the independence assumption which is not given for
  overlapping segments. The test is applicable only for two
  independent time series and would suffer from the multiple
  comparison problem for multiple segments.  \citet{dette:2007} use a
frequency-domain bootstrap based on the $ L_2$ between two
nonparametrically estimated PSDs and pooled PSD. It
  does not make the assumption of independence but requires the
  estimation of the spectral density matrix which would only be
  possible with considerable computational time in the case of
  spectrograms. In general, the power of bootstrap tests for
  stationarity depends on the particular type of bootstrap and though
  asymptotically consistent under certain conditions, they do not
  provide general finite-sample guarantees
  \citep{McCulloughMegan2013Tsww}.

To avoid deficiencies of the bootstrap methods, our
tests fall into the lesser-known \textit{surrogate data} tests which
were first introduced by \citet{theiler:1992} for testing
non-linearities in time series, and later adapted by \citet{xiao:2007}
and \citet{borgnat:2009} for testing stationarity.  These tests are
nonparametric in nature, where the original data are resampled to
create stationary surrogates with the same periodogram.  A version of
the multitaper spectrogram of \citet{thomson:1982} with Hermite
(rather than Slepian) window functions (as discussed by
\citet{bayram:2000}) is computed, where the estimated spectrum in each
time segment is compared to a time-averaged spectrum using a distance
measure, typically a combination of the Kullback-Leibler divergence
and the log spectral deviation.  The test statistic for these tests
are the sample variance of these distances and a 
  Gamma distribution is fitted to describe the null distribution of
  test statistics.

In this paper, we propose two variants on the surrogate data testing
of \citet{xiao:2007} and \citet{borgnat:2009} that do
  not rely on the Gamma distributon to describe the distribution of
  the test statistic under the null hypothesis.  We consider an
autoregressive spectrogram where each short-time segment uses a
frequentist autoregressive (AR) estimate of its spectrum, with order
selected based on the Akaike information criterion (AIC).  In the
first variant, we can compute the Kolmogorov-Smirnov statistic, the
Kullback-Leibler distance, or the log spectral
  distance to measure the distance between local spectra of short
time segments and the global spectrum.  A test statistic is then
computed as the sample variance of these distances and we use
surrogates to populate the sampling distribution of this test
statistic under the null hypothesis of stationarity.  Large
variability in the distances of the original time series would provide
evidence against stationarity.  As a novel alternative, we fit a least
squares regression line to the cumulative median of Euclidean
distances between columns in the AR spectrogram.  The slope of this
line is used as a test statistic and surrogates are again used to
generate the null distribution.  Here, if a time series is stationary,
we would expect the PSD in neighbouring segments of the spectrogram to
be similar over time, meaning the median of Euclidean distances should
fluctuate around a constant.  A non-zero slope would then provide
evidence against the stationarity hypothesis.  In both variants,
empirical percentiles are used to create a critical value that is used
as a rejection threshold.

We introduce these hypothesis tests to be used as a tool for future
LISA data analysis, with the overall goal of determining how often we
should update the noise PSD.  Once this is decided, parameter
estimation routines can be implemented.  In this paper, we propose the
use of a blocked Metropolis-within-Gibbs sampler to simultaneously
estimate the parameters of a galactic white-dwarf binary gravitational
wave signal and estimating the noise PSDs before and after a planned
data gap.  We show that the stationarity tests based on the surrogate data approach
can be applied to the residuals to check the validity of model assumptions.


The paper is structured as follows.  In Section~\ref{sec:identifying},
we introduce the notion of surrogate data testing, defining two
specific hypothesis tests to be used in the future LISA mission.  We
then conduct a simulation study to demonstrate the power of these
tests, and then apply the tests to differential acceleration
measurements from LPF to highlight nonstationarities in that data.  In
Section~\ref{sec:addressing}, we introduce our data analysis strategy
for handling nonstationary LISA noise.  We inject a galactic
white-dwarf binary GW signal in piecewise stationary noise and
implement a blocked Metropolis-within-Gibbs sampler
for posterior computation of both signal parameters
  and noise PSDs.  We mimic what we believe could happen to LISA
noise when repointing satellites during planned gaps,
and apply stationarity tests to residuals for model
  checking.
 We then give concluding remarks in
Section~\ref{sec:discussion}.

\section{Identifying Nonstationary Noise}\label{sec:identifying}

\subsection{Stationary Surrogates}\label{sec:surrogate}

Surrogate data testing was originally proposed by \citet{theiler:1992}
for testing non-linearities in time series, and later adapted by
\citet{xiao:2007} and \citet{borgnat:2009} for testing stationarity.
The main idea here is that one can create stationary ``surrogates'' of
a (potentially nonstationary) time series   by directly manipulating the
data in the frequency-domain, preserving the second-order statistics,
but randomizing higher order statistics.  In this way, we can generate
a stationary surrogate of a time series that has the same empirical
spectrum (periodogram) as the original time series.

First, Fourier transform the time series  $Y(t), t=1,\ldots,n$ using
\begin{equation*}
\tilde{Y}(\omega_j) = \sum_{t=1}^{n} Y(t) \mathrm{e}^{-\mathrm{i} t \omega_j}
\end{equation*}
to get a frequency-domain representation where 
$\omega_j=2\pi j/n, j=0,\ldots,n-1$, 
are the Fourier frequencies. The Fourier coefficients can be  expressed  in polar
coordinates such that
\begin{equation*}
  \tilde{Y}(\omega_j) = A(\omega_j)\mathrm{e}^{\mathrm{i}\varphi(\omega_j)},
\end{equation*}
where $A(\omega_j) = |\tilde{Y}(\omega_j)|$ is the magnitude vector and
$\varphi(\omega_j) = \arg\left(\tilde{Y}\left(\omega_j\right)\right)$ is the
phase vector.

Keeping the magnitude vector $(A(\omega_0),\ldots,A(\omega_{n-1}))$ fixed, we replace the phase
vector $(\varphi(\omega_1),\ldots,\varphi(\omega_{n-1}))$ by a new phase vector $(\varphi^*(\omega_1),\ldots,\varphi^*(\omega_{n-1}))$
that is populated by iid $\mathrm{Uniform}[0, 2\pi]$ random
variables.  We now have a randomized frequency-domain representation
of the surrogate $\tilde{Y}^*(\omega_j) = A(\omega_j)
\mathrm{e}^{\mathrm{i}\varphi^*(\omega_j)}$ which is inverse Fourier
transformed to give a time-domain representation of the surrogate:
\begin{equation*}
Y^*(t) = \frac{1}{n} \sum_{j=0}^{n-1} \tilde{Y}^*(\omega_j) \mathrm{e}^{\mathrm{i} t \omega_j} .
\end{equation*}
Assume $n$ is even and 
let $(\omega_0, \omega_1, \ldots, \omega_{n / 2 - 1}, \omega_{n /
  2})$ be the first Fourier frequencies.  We only randomize the
phase for $\omega_1, \omega_2, \ldots, \omega_{n/2-1}$ because
$\omega_0$ and $\omega_{n / 2}$ are always real-valued with zero
phase, and the subsequent $n/2$ Fourier coefficients are complex conjugates of
the first Fourier coefficients for the inverse Fourier transform to
	be real-valued, meaning $\varphi(\omega_j) = -\varphi(\omega_{n-j})$.

Surrogates are extremely useful for testing stationarity as they not
only have the same periodogram as the original data (which may or may
not be stationary), but they are stationary themselves, meaning if one
can compute a test statistic that can distinguish the null hypothesis
(stationary) from the alternative hypothesis (nonstationary), it is
straightforward to generate the sampling distribution of the test
statistic by computing the test statistic on a large number of
surrogates.  We now focus our attention on useful test statistics
based on the autoregressive spectrogram.

\subsection{Autoregressive Spectrogram}\label{sec:ar_spectro}

The spectrogram is the most fundamental tool used in time-frequency
analysis. It contains at each column an approximation of the PSD
function for consecutive time intervals.  Thus, it allows us to assess
the evolution of this function over time.  It is computed as
follows. First compute the short-time Fourier transform (STFT),
\begin{equation*}
\tilde{Y}(\omega, T) = \int W(t-T)Y(t) \mathrm{e}^{-\mathrm{i}t\omega} dt,
\end{equation*}
where $W(.)$ is a window function of duration $T$.  Then take the
squared modulus of each segment.  This amounts to computing the
periodogram of short windowed segments of the data, which may or may
not be overlapping in time.

It is well-known in the time series literature that the periodogram is
an asymptotically unbiased estimator of the spectral density function,
but it is not a consistent estimator.  This has lead to a large amount
of literature on periodogram smoothing to reduce the variance.

The most popular parametric approach is to fit an autoregressive model
where the order chosen by AIC.  In this paper, we use an AR estimate
of the spectrum for each segment of the spectrogram rather than using
the raw periodogram.  Although there are more sophisticated approaches
to spectrum estimation that perhaps do not rely on parametric
assumptions (see for example \citet{choudhuri:2004},
\citet{edwards:2017}, \citet{kirch:2018}, \citet{russel:2019b} for
novel Bayesian approaches), we use the frequentist AR method for the
sake of computational speed and ease.

For the remainder of the paper, when computing the AR spectrogram, we
utilize the Tukey window with tapering coefficient equal to $(1 -
\mathrm{Overlap}) / 10$, where $\mathrm{Overlap}$ is
  the proportion of data that neighbouring time segments coincide.


\subsection{Variance of Local Contrast (VOCAL) Test}\label{sec:vocal}

In this section, we describe the first of two surrogate tests, which
we call the Variance of Local Contrast (VOCAL) Test.  As with any
hypothesis test, we need to first define a test statistic that can
distinguish between the null hypothesis and alternative hypothesis.

First consider the original time series and find its AR spectrogram.
We need to contrast local features in the spectrogram with the global
spectrum by computing a \textit{local contrast} for each time segment
(column) in the spectrogram.  This is computed as
\begin{equation*}
c_l = \kappa(\hat{f}_l, \hat{f}), \quad l = 1, 2, \ldots, L,
\end{equation*}
where $L$ is the number of time segments (columns) in the spectrogram,
$\hat{f}_l$ is the estimated (local) PSD of the $l^{\mathrm{th}}$ time
segment of the spectrogram, $\hat{f}$ is the estimated (global) PSD of
the entire time series (estimated using the same AR routine in the
spectrogram), and $\kappa$ is a suitable spectral distance,

In this paper we use three different distance or
  dissimilarity measures $\kappa$ to specify the local contrasts.  The
  first one uses the Kolmogorov-Smirnov (KS) statistic
\begin{equation*}
\kappa^{(1)}(f_1, f_2) = \sup_{\omega} |F_1(\omega) - F_2(\omega)|,
\end{equation*}
where $F_1$ and $F_2$ are standardized empirical cumulative
distribution functions (ECDFs) computed by normalizing the estimated
PSDs $f_1$ and $f_2$ (such that they integrate to 1
  and can be considered to be probability density functions), and
taking their cumulative sums.  The second one uses the symmetric
Kullback-Leibler (KL) divergence
\begin{equation*}
\kappa^{(2)}(f_1, f_2) = \frac{1}{2} \int \Big(f_1\left(\omega\right) -
f_2\left(\omega\right) \Big) \log
\frac{f_1(\omega)}{f_2(\omega)}\mathrm{d}\omega,
\end{equation*}
where $f_1$ and $f_2$ are normalized PSDs. The third
  is the log spectral distance (LSD), a dissimilarity measure defined
  directly on the unnormalized spectral densities by
\begin{equation*}
\kappa^{(3)}(f_1, f_2) =  \int  \left|\log \frac{f_1(\omega)}{f_2(\omega)} \right| d\omega.
\end{equation*}
Whereas the KS and KL distance are insensitive to any changes in scale
of the PSD because of the normalization, the LSD is well suited to
quantify differences in both shape and scale such as amplitude
modulations.

Fluctuations in the local contrasts can be used to distinguish between
stationarity and nonstationarity as we would expect very little
variability in the local contrasts if a time series was stationary and
more variability if the time series was nonstationary.  To this end,
we use the sample variance of local contrasts as the test statistic
for this test, i.e.,
\begin{equation*}
  V = \mathrm{Var}(\mathbf{c}),
\end{equation*}
where $\mathbf{c} = (c_1, c_2, \ldots, c_L)$.

We can then generate the sampling distribution of this test statistic
under the null hypothesis by repeating this same process on stationary
surrogate data.  That is, for each surrogate (indexed by $s = 1, 2,
\ldots, S$, for large $S$) compute the AR spectrogram, the local
contrasts $\mathbf{c}_s$, and finally the test statistic to give us
\begin{equation*}
  V_0(s) = \mathrm{Var}(\mathbf{c}_s), \quad s = 1, 2, \ldots, S,
\end{equation*}
where $\mathbf{c}_s = (c_{s, 1}, c_{s,2}, \ldots, c_{s, L})$.

The hypothesis test can then be formalized by considering where $V$
lies in the distribution of $V_0$.  Let
\begin{eqnarray*}
  H_0&:& V < \gamma \quad \mathrm{(Stationary)},\\
  H_1&:& V \geq \gamma \quad \mathrm{(Nonstationary)},
\end{eqnarray*}
where $\gamma$ is the critical value chosen such that
\begin{equation*}
  p(V_0 \leq \gamma) = 1 - \alpha,
\end{equation*}
where $\alpha$ is the rejection threshold.  Thus for an $\alpha =
0.05$ rejection threshold, $\gamma$ is computed as the 95\% percentile
of $V_0$.  Alternatively, an approximate $p$-value can be computed by
\begin{equation*}
  \frac{1}{S}\sum_{s = 1}^S I_{\{V_0(s) \geq V\}},
\end{equation*}
where $I$ is an indicator function.  Note that this is a one-sided
test.

The precision to which the $p$-value can be computed depends on the
number of surrogates generated.  For example, if $S = 1,000$, the
$p$-value can be computed to three decimal places, and if $S =
10,000$, the $p$-value can be computed to four decimal places.

As an illustrative example of the test, consider the autoregressive
(AR) model, defined as:
\begin{equation*}
  Y_t = \sum_{i = 1}^p \varphi_i Y_{t-i} + \varepsilon_t,
\end{equation*}
where $p$ is the order, $(\varphi_1, \ldots, \varphi_p)$ are the model
parameters, and $\varepsilon_t \sim \mathrm{N}(0, \sigma^2)$ for all
$t$ is the white noise innovation process.

\begin{figure}[!h]
\includegraphics[width=1\linewidth]{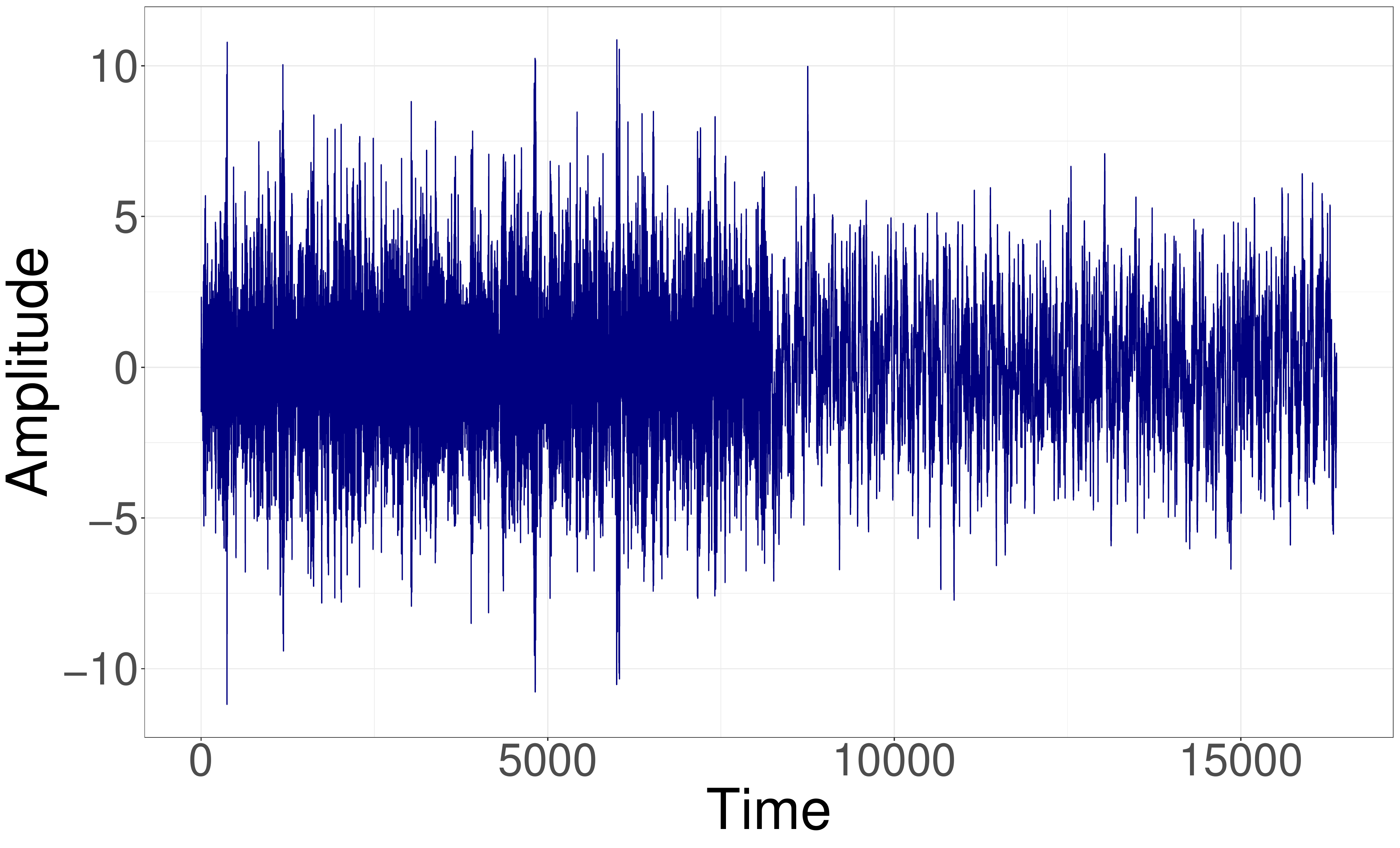}
\caption{Time series containing $2^{13}$ realizations from an AR(2)
  with parameters $(0.9, -0.9)$ and $2^{13}$ realizations from an
  AR(1) with parameter 0.9.  Each series uses N$(0, 1)$ innovations.}
\label{fig:Example_Data}
\end{figure}

Consider the case where we have a length $n = 2^{13}$ time series
generated from an AR(2) with parameters (0.9, -0.9), and we
concatenate this with a length $n = 2^{13}$ time series generated from
an AR(1) with parameter 0.9, each with standard normal innovations, as
illustrated in Figure~\ref{fig:Example_Data}.

\begin{figure}[!h]
\includegraphics[width=1\linewidth]{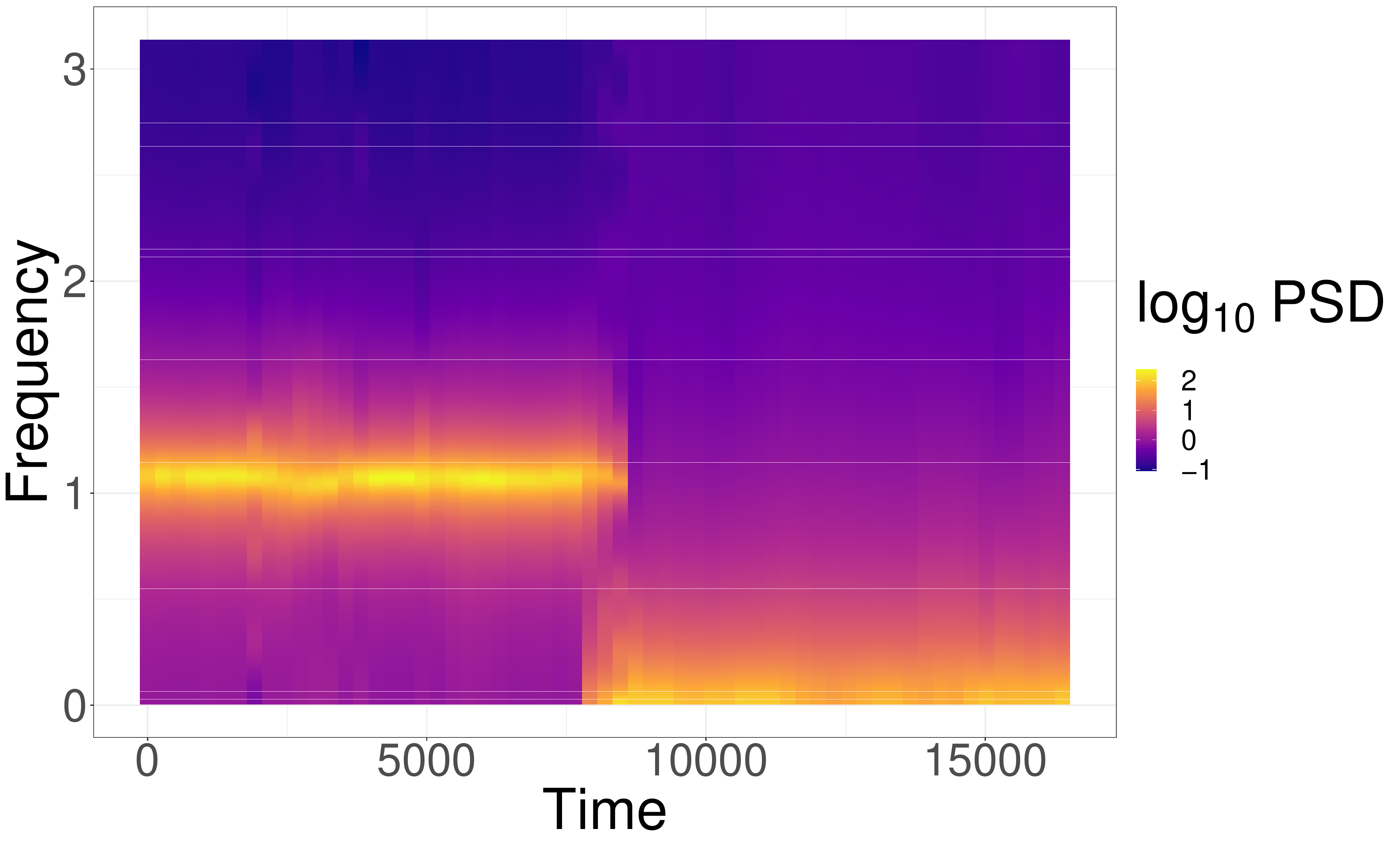}
\caption{AR spectrogram from the time series presented in
  Figure~\ref{fig:Example_Data}.  Notice the abrupt change in PSD
  structure at the halfway point.}
\label{fig:Example_Spectrogram}
\end{figure}

Setting the overlap to 75\% and window length to $2^{10}$, the
associated AR spectrogram can be seen in
Figure~\ref{fig:Example_Spectrogram}.  Notice how the spectrum changes
around halfway through the time series.

\begin{figure}[!h]
\includegraphics[width=1\linewidth]{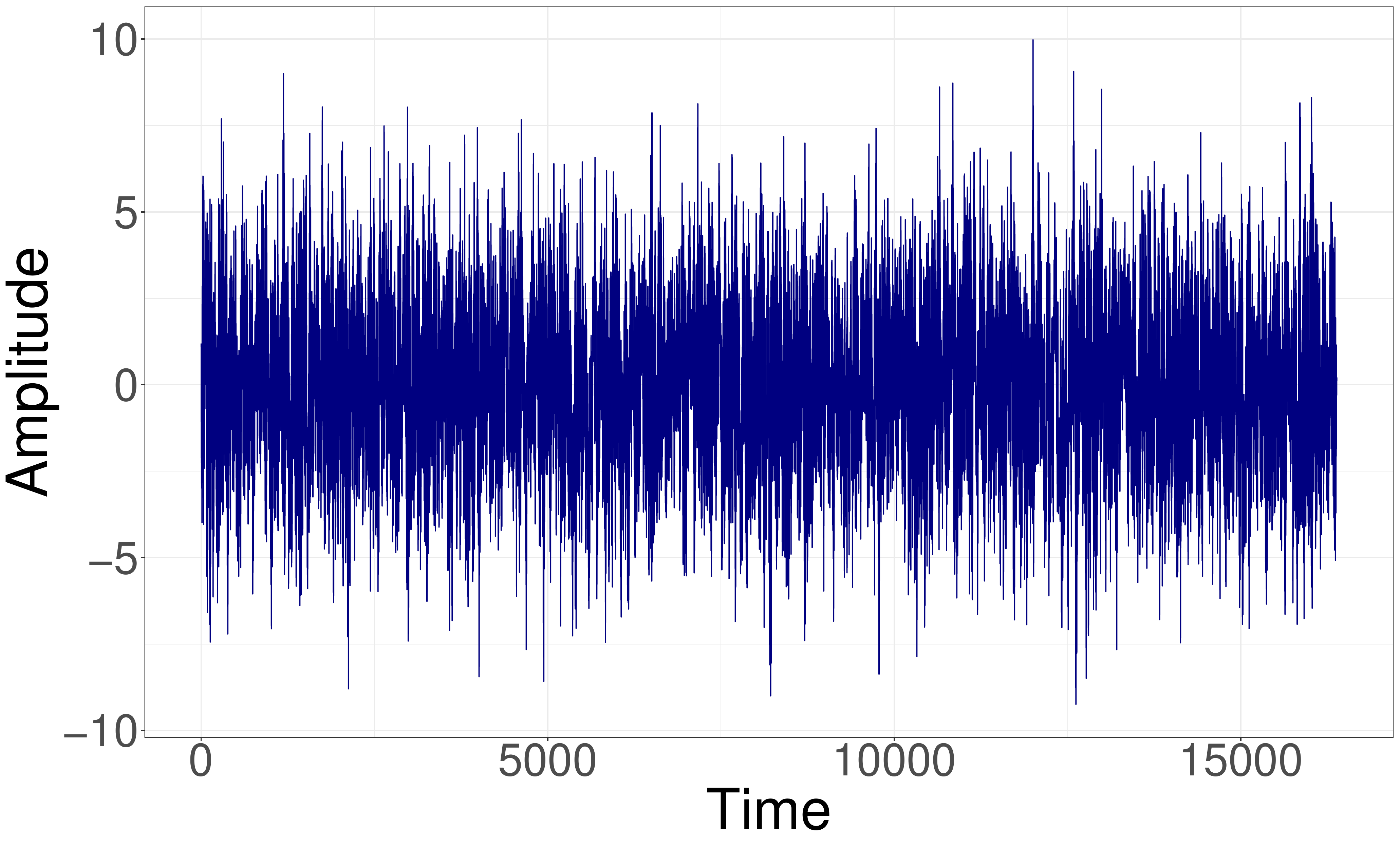}
\caption{One example of stationary surrogate data based on the time
  series presented in Figure~\ref{fig:Example_Data}.}
\label{fig:Surrogate_Data}
\end{figure}

\begin{figure}[!h]
\includegraphics[width=1\linewidth]{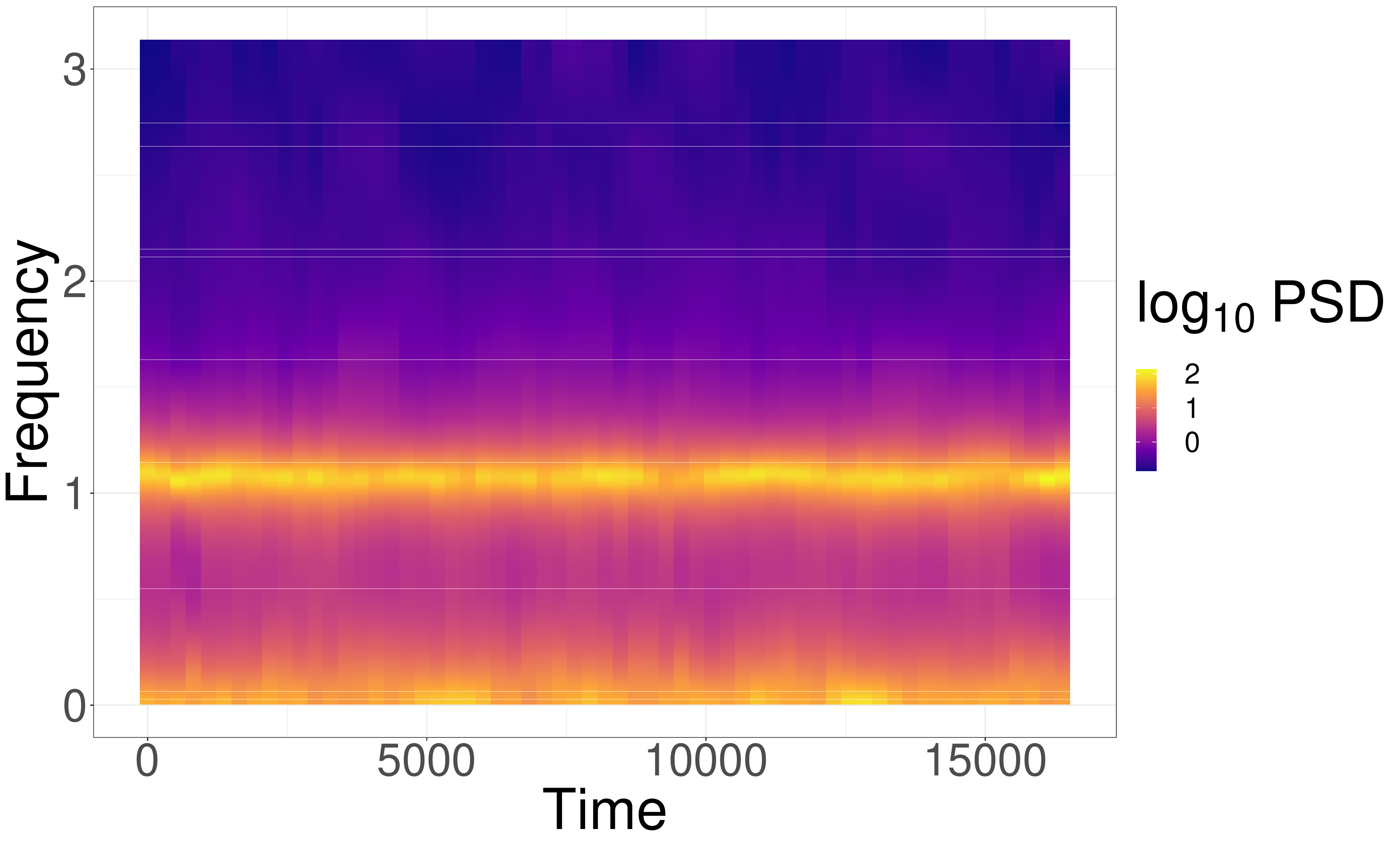}
\caption{AR spectrogram from the stationary surrogate data presented
  in Figure~\ref{fig:Surrogate_Data}.}
\label{fig:Surrogate_Spectrogram}
\end{figure}

We now generate 1,000 surrogates.  One example of a surrogate of our
original time series can be seen in Figure~\ref{fig:Surrogate_Data}
and its associated AR spectrogram can be seen in
Figure~\ref{fig:Surrogate_Spectrogram}.

\begin{figure}[!h]
\includegraphics[width=1\linewidth]{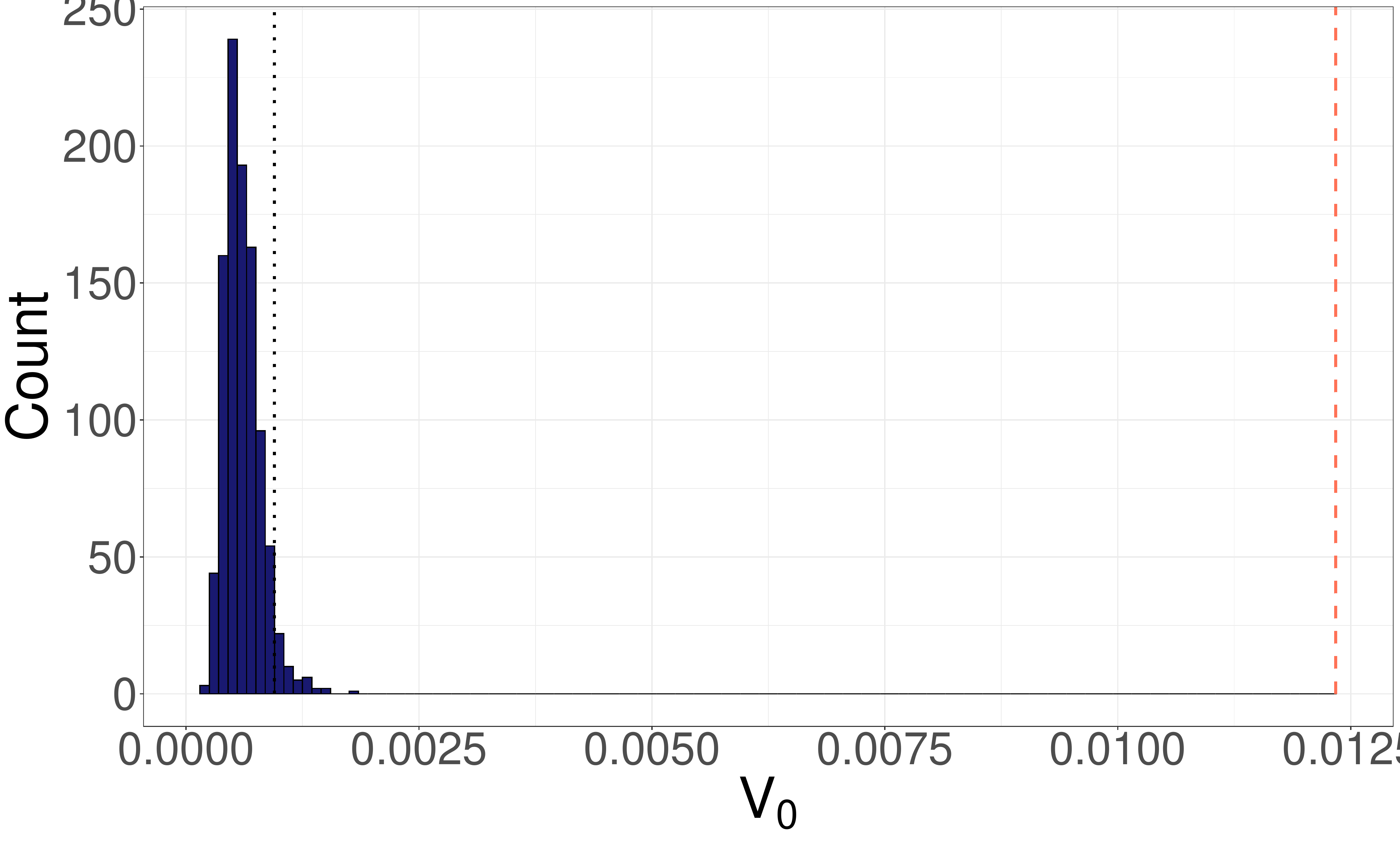}
\caption{Empirical sampling distribution of test statistic (variance
  of local contrasts computed using the KS statistic).  The dotted
  black line is $\gamma$ (the 95\% percentile of this null
  distribution) and the dashed pink line is the test statistic $V$
  from the original time series.}
\label{fig:VOCAL_KS_Histogram}
\end{figure}

Using the KS statistic as the local contrast, we can generate the test
statistic $V$ from the original data, and the empirical sampling
distribution of the test statistic using $(V_0(1), V_0(2), \ldots,
V_0(S))$.  Using a 5\% rejection threshold, we compute the 95\%
percentile of the empirical sampling distribution.  This is
illustrated in Figure~\ref{fig:VOCAL_KS_Histogram}.  As the test
statistic $V$ is greater than the 95\% percentile of the empirical
sampling distribution, we reject the null hypothesis of stationarity.

\subsection{Slope of Median Euclidean Distance (SOMED) Test}\label{sec:somed}

For our second surrogate
test, we compare the \textit{Euclidean distances} between the
estimated PSD functions over time, i.e., a comparison between the
columns of the spectrogram.  If a time series is stationary, each column 
in the spectrogram should look approximately similar over time (see e.g.,
Figure~\ref{fig:Surrogate_Spectrogram}).  Consequently, a
sequence of consecutive distances should fluctuate around a constant.
We propose to test stationarity by testing the significance of the
slope in a simple linear regression model fitted to these distances.

First, we calculate the AR spectrogram.  This conforms a matrix $(r
\times m)$ where the rows and columns stand for the energy or power at
a particular frequency and the time intervals, respectively.  Then, we
calculate the Euclidean distance of each column with respect to the
other ones, that is
\begin{align*}
	d_{ij} = \sqrt{\sum_{k=1}^{r} \left(Y_{ki} - Y_{kj}\right)^2},
\end{align*}
where $\textbf{Y}_i = (Y_{1i}, \dots, Y_{ki}, \dots, Y_{ri})^\top$ is
the $i^{\text{th}}$ column of the spectrogram for $i=1,\dots,m$.  The
distances $d$ compound a symmetric matrix $\textbf{D}$ which has a
vector of zeros in its diagonal.

Since $\textbf{D}$ is symmetric, we discard the upper triangular part
and calculate the median of each row, which generates a sequence
$\textbf{v} = (v_2, \dots, v_m)$, where $v_i$ is the median of the
Euclidean distances of the estimated PSD for the $i^{\text{th}}$ time
interval (column in the spectrogram matrix) with respect to all the
estimated PSD of the previous time intervals, i.e., it is a cumulative
median.  Since the first $v_i$ values embody a few comparisons that
tend to generate low discrepancies, these can be discarded, for
instance, the first 10\% of the sequence.

If the time series is stationary, we would expect a similar PSD across
time.  In other words, the cumulative median of the Euclidean
distances should fluctuate around a constant, which can be tested
evaluating the slope of a fitted simple linear regression model.
Thus, we fit a linear model $y_i = \beta_0 + \beta_1 x_i +
\varepsilon_i$, where the responses are the sequence $\textbf{v}$ and
the explanatory variables points in time.  We assume that the errors
$\varepsilon_i$ are independent and identically distributed with
$\mathbb{E}(\varepsilon_i)=0$ and $\text{Var}(\varepsilon_i) =
\sigma^2$.  If the estimated slope is zero it means that the time
series is stationary, otherwise the time series is nonstationary.  We
assess this assumption of the time series through the following
hypotheses:
\begin{align*}
H_0&:\beta_1 = 0 \quad \text{(Stationary)}\\
H_1&:\beta_1 \neq 0 \quad \text{(Nonstationary)}.
\end{align*}
The null hypothesis establishes that the sequence of medians
$\textbf{v}$ does not change over time or equivalently the PSD
functions do not vary significantly over time, showing the
stationarity of the time series.

To test $H_0$, we compare the slope estimated from the original data
$\widehat{\beta}$ with the empirical distribution of the slopes
estimated from surrogate data sets $\bmt{\widehat{\beta}}_S =
(\widehat{\beta}_1,\dots,\widehat{\beta}_S)$, i.e., under the null
hypothesis that assumes stationarity.  Then, the $p$-value is calculated
by
\begin{equation*}
\dfrac{1}{S}\sum_{s=1}^{S}\left( I_{\left\{-\left|\widehat{\beta}\right| > \widehat{\beta}_s\right\}} +
I_{\left\{\left|\widehat{\beta}\right| < \widehat{\beta}_s\right\}} \right),
\end{equation*}
where $I$ is an indicator function.

This test also has the potential of detecting glitches using
conventional statistical techniques used to detect outliers in linear
regression models.  This can be assessed by analyzing the cumulative median 
values of the original data set.

\begin{figure}[!h]
	\includegraphics[width=1\linewidth]{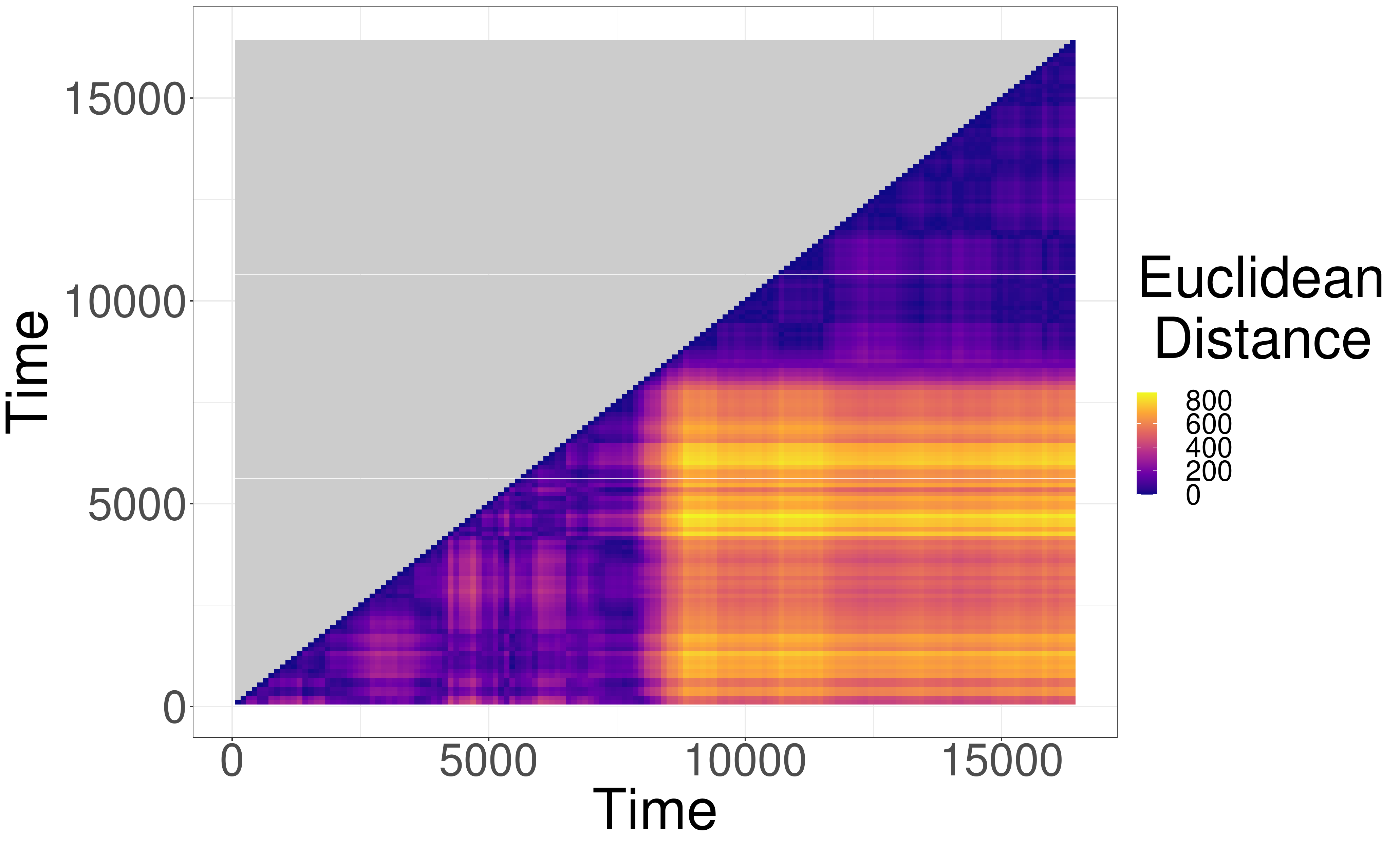}
	\caption{Euclidean distances for the spectrogram displayed in
          Figure~\ref{fig:Example_Spectrogram}.}
	\label{fig:Example_Euclidean}
\end{figure}

Consider the AR spectrogram used in Section~\ref{sec:vocal}.  The
nonstationary design of this process can be clearly noted in the
spectrogram displayed in Figure~\ref{fig:Example_Spectrogram}.  The
two PSDs corresponding to the AR(2) and AR(1) processes have their
peaks at different frequencies.  This difference is also clear in the
comparison of the Euclidean distances displayed in
Figure~\ref{fig:Example_Euclidean}.  The discrepancy in the PSD
estimates is represented in the magnitude of the distances which
conform a block in the lower-right part.


\begin{figure}[!h]
	\includegraphics[width=1\linewidth]{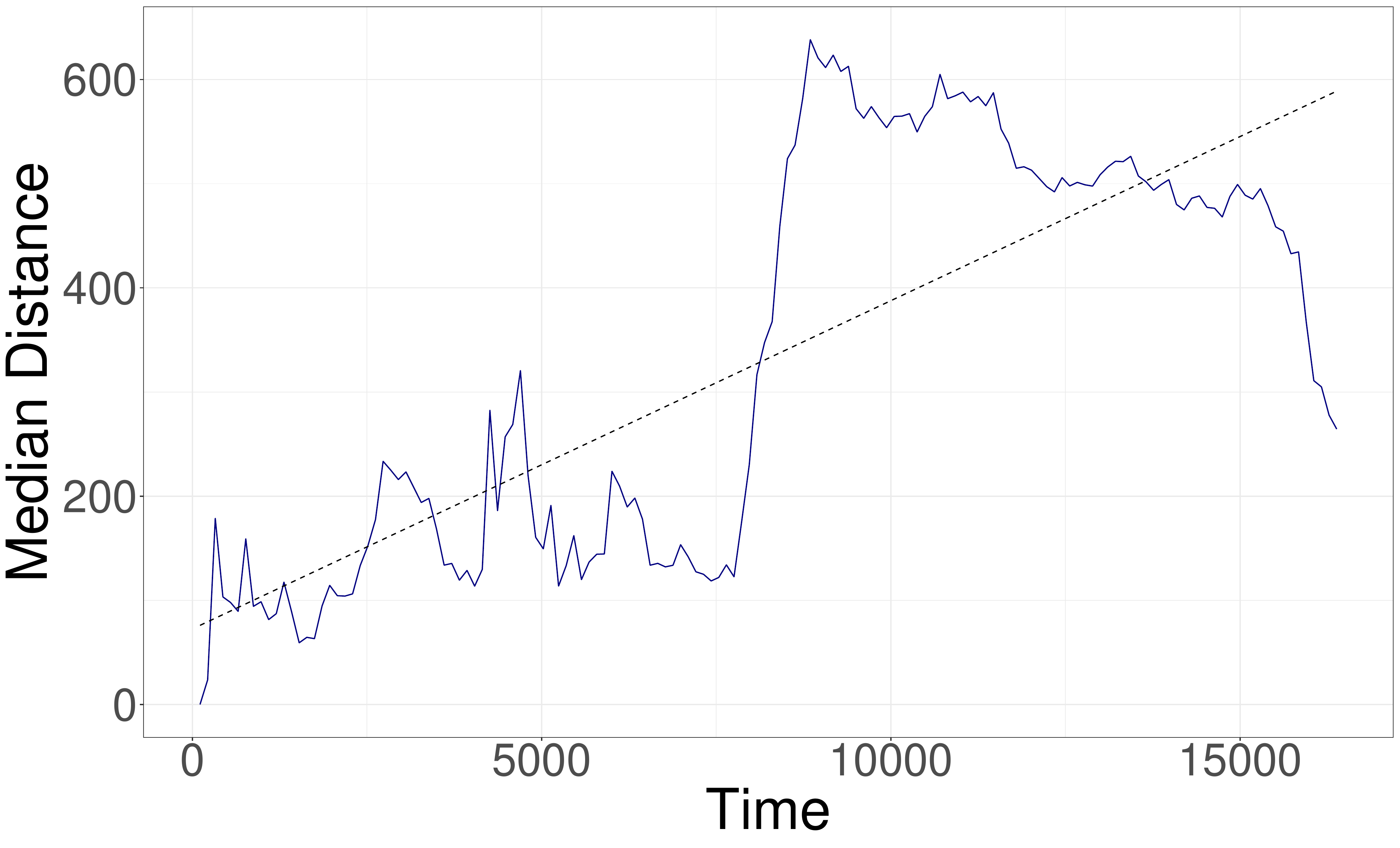}
	\caption{Median of the Euclidean distances for each column of
          Figure~\ref{fig:Example_Euclidean}.  The dashed line stands
          for a simple linear model.}
	\label{fig:Example_Median}
\end{figure}

The medians of the Euclidean distances of a specific time interval in
Figure~\ref{fig:Example_Euclidean} with respect to its previous
intervals are displayed in Figure~\ref{fig:Example_Median}.  It can be
noticed the design of the process: the first half is centred below the
second one.  The slope of the simple linear model is evidently non
zero. The discrepancy of the PSD estimates do not seem to fluctuate
randomly around a constant, which is evidence in favour of the
nonstationary nature of the process.  Comparing this slope with the
empirical distribution of the slopes calculated from the surrogate
data sets we get a $p$-value of 0.000.  The SOMED test rejects the
null hypothesis, identifying successfully this data set as
nonstationary.

\subsection{Testing Simulated Data}\label{sec:testing}

We now apply the surrogate tests to simulated AR data (with standard
white noise innovations) and compute power or size for different
scenarios.  Consider a length $n = 2^{12}$ time series $\mathbf{Y}$
that is split in half into two length $n / 2 = 2^{11}$ time series
$\mathbf{Y}_1$ and $\mathbf{Y}_2$.  For the following three scenarios,
let $\mathbf{Y}_1$ and $\mathbf{Y}_2$ have the:
\begin{enumerate}
\item Same dependence structure;
\item Different dependence structure;
\item Similar dependence structure;
\end{enumerate}
where ``dependence structure'' refers to the
  autocovariance function of a time series, or equivalently the
  spectral density function, which is its Fourier transform.

In Scenario 1, we consider a time series with the same dependence
structure (and therefore same PSD) throughout its duration.  Let
$\mathbf{Y}_1$ and $\mathbf{Y}_2$ be generated from an AR(1) with
parameter 0.9.  In this scenario, we show that both tests yield small
Type I Errors, i.e.\ do not reject the null hypothesis of stationarity
the vast majority of times.

In Scenario 2, we look at an extreme example, where $\mathbf{Y}_1$ and
$\mathbf{Y}_2$ have vastly different dependence structures.  Let
$\mathbf{Y}_1$ be generated from an AR(2) with parameters (0.9, -0.9)
and $\mathbf{Y}_2$ be generated from an AR(1) with parameter 0.9.
Here, we demonstrate that both methods reject the null hypothesis of
stationarity, with high power.

In Scenario 3, we let $\mathbf{Y}_1$ and $\mathbf{Y}_2$ have very
similar (but not equivalent) dependence structures.  Let
$\mathbf{Y}_1$ come from an AR(1) with parameter 0.8 and
$\mathbf{Y}_2$ come from an AR(1) with parameter 0.9. 

Finally we add a fourth scenario:
\begin{enumerate}
  \setcounter{enumi}{3}
\item Time-varying dependence structure.
\end{enumerate}
We use a time-varying autoregressive model (TVAR), where coefficients
vary linearly from -0.6 to 0.6.  Here, we demonstrate that both
approaches reject the stationarity hypothesis when the spectrum is
time-varying, with high power.\\

For each scenario we generate a time series, compute its AR
spectrogram, and test statistic.  We then create 1,000 stationary
surrogates, compute their AR spectrograms and test statistics and
compare the observed test statistic against the sampling distribution
of test statistics.  If the observed test statistic is in the tails of
the distribution, this gives us evidence against the stationarity
hypothesis.  Specifically, we use the 95\% percentile as the critical
value for the one-sided VOCAL tests (i.e., a $p$-value of $< 0.05$),
and $p$-value of $< 0.05$ for the two-sided SOMED test.

The AR spectrograms are generated using a window length of $T = 2^9$,
and overlap of 75\%.  We conduct both the VOCAL and the SOMED
hypothesis tests, and consider the KS, KL, and LSD
variants on the VOCAL test.

We replicate each simulation 1,000 times and report the size or power
of each test, at the 5\% significance level, where the size of a test
is the probability of falsely rejecting the null hypothesis when it is
true (or the probability of making a Type I Error), and the power of a
test is the probability of correctly rejecting the null hypothesis
when it is false (or one minus the probability of making a Type II
Error).  Type I and II Errors are equivalent to
  \textit{false positives} and \textit{false negatives} respectively.
Our results are presented in Table~\ref{tab:power}.

\begin{table}[!h]
    \begin{center}
      \caption{\label{tab:power} Test size (probability of falsely
        rejecting $H_0$ when it is true) for Scenario 1, and test
        power (probability of correctly rejecting $H_0$ when it is
        false) for Scenarios 2, 3, and 4.}
    \begin{tabular}{ccccc}
      \hline
    Scenario&KS&KL&LSD&SOMED\\
    \hline
    \hline
    1&0.036&0.048&0.046&0.046\\
    \hline
    \hline
    2&1.000&1.000&1.000&1.000\\
    3&0.794&0.739&0.049&0.962\\
    4&1.000&1.000&1.000&0.999\\
    \hline
    \end{tabular}
  \end{center}
\end{table}

We see that when $\mathbf{Y}_1$ and $\mathbf{Y}_2$ have the same PSD,
all tests have a very small test size and that there is less than a
5\% chance of making a Type I error.  For the extreme case where
$\mathbf{Y}_1$ and $\mathbf{Y}_2$ have very different PSDs, all tests
give us power 1, which means there is zero chance of making a Type II
error.  In the case where we have similar but not equivalent PSDs, all
tests reject the null hypothesis the majority of the time and the
SOMED test works particularly well, which is remarkable considering
how similar the $\mathbf{Y}_1$ and $\mathbf{Y}_2$
are. The LSD test, though, has very low power in this
  scenario, as it is less suited to discriminate between small changes
  in distributional shapes than the KL and KS distance measures. When
we have a time-varying PSD, we again have high power.  All of these
results give us great confidence that the surrogate tests are
performing as required.

\subsection{LISA Pathfinder}\label{sec:lpf}

We now demonstrate that our surrogate tests can detect
nonstationarities in the clean (Level 3) $\Delta g$ data from the
noise runs of LPF.  These data have been corrected for the
acceleration coming from centrifugal force, acceleration on the $x$-axis
coming from the spacecraft motion along other degrees of freedom, and
spurious acceleration noise from the digital to analog converter of
the capacitive actuation and Euler force.  Details can be found in the
technical note on the LPF data archive \citep{archive}.

We analyze segments from two separate noise runs.  These have the
following starting times and lengths:
\begin{enumerate}
\item 2016-04-03 14:55:00 UTC for 12 days, 16 hours, 29 minutes, 59.40
  seconds.  We refer to this data set as the \textit{Glitch Data Set}.
\item 2017-02-13 07:55:00 UTC for 18 days, 13 hours, 59 minutes, 59.40
  seconds.  We refer to this data set as the \textit{Amplitude
    Modulation (AM) Data Set}.
\end{enumerate}

The LPF data are originally sampled at a rate of 10~Hz (with sample
interval $\Delta_t = 0.1$~s).  For the Glitch Data Set, we downsampled
the data to 0.2~Hz ($\Delta_t = 5$~s) to obtain a Nyquist frequency of
0.1~Hz (but first Tukey windowing with parameter 0.01, then applying a
low-pass Butterworth filter of order 4 and critical frequency 0.1~Hz
to avoid aliasing issues). The frequency range of interest for most GW
signals detectable by LISA is $[10^{-4}, 10^{-1}]$~Hz.  To resolve the
lowest frequency in this band, the shortest (base 2) time series we
can analyze is $n = 2^{11}$.  We therefore split the data into
non-overlapping segments of length $n = 2^{11}$ to speed up
computations.

\begin{figure}[!h]
\includegraphics[width=1\linewidth]{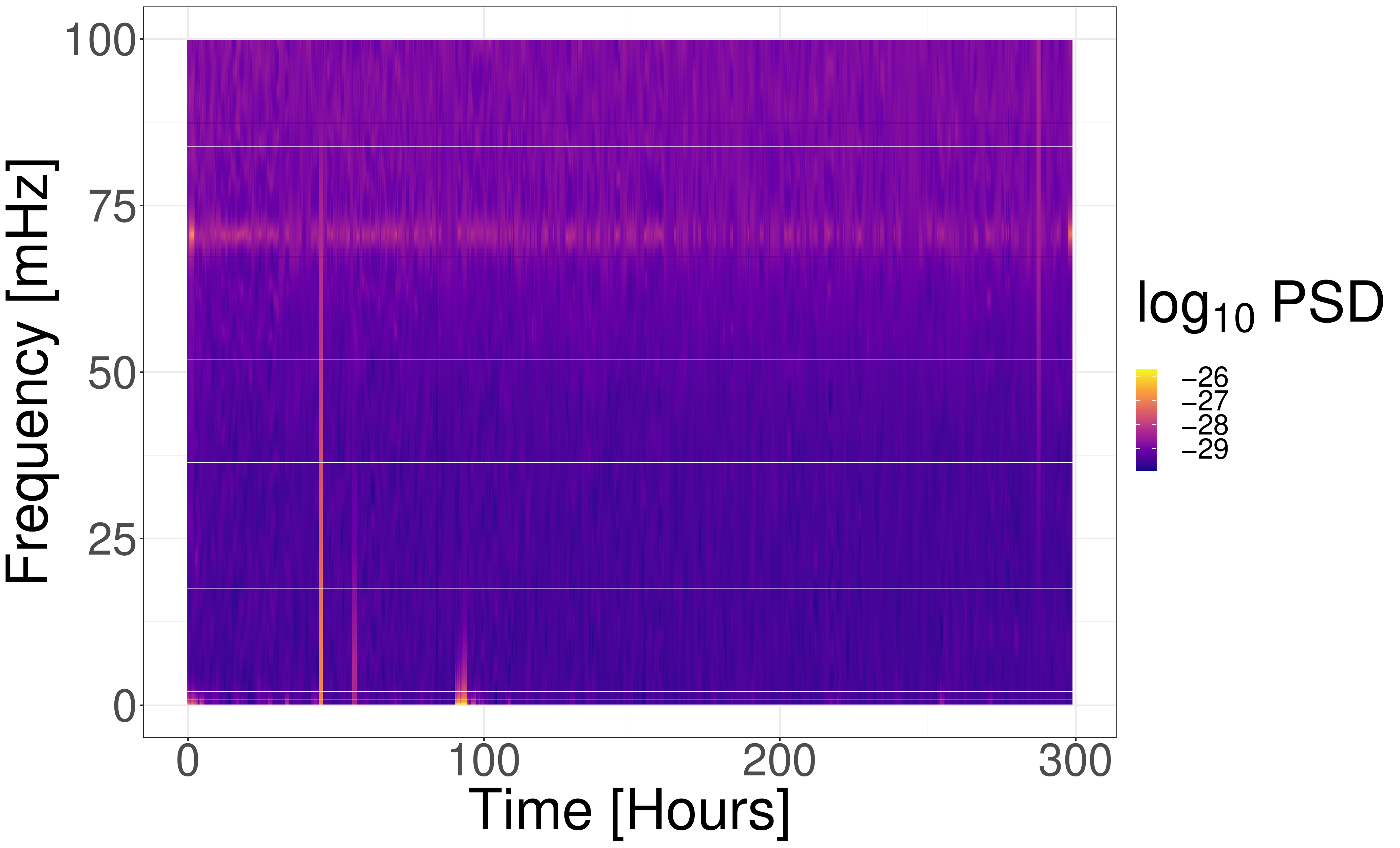}
\caption{AR Spectrogram of the Glitch Data Set.}
\label{fig:LPF_Glitch_Spectrogram}
\end{figure}

It is important to note that in the mean sense of stationarity, once
filtered and downsampled, the Glitch Data Set is nonstationary, as
there is a trend.  We therefore remove this trend piecewise linearly
for each non-overlapping segment, and we focus our attention on the
question of whether LPF noise is nonstationary in terms of its
autocovariance function, or equivalently its PSD.  The
  AR spectrogram (with window length $T = 2^{10}$ and 75\% overlap) of
  the Glitch Data Set can be seen in
  Figure~\ref{fig:LPF_Glitch_Spectrogram}.

\begin{figure}[!h]
\includegraphics[width=1\linewidth]{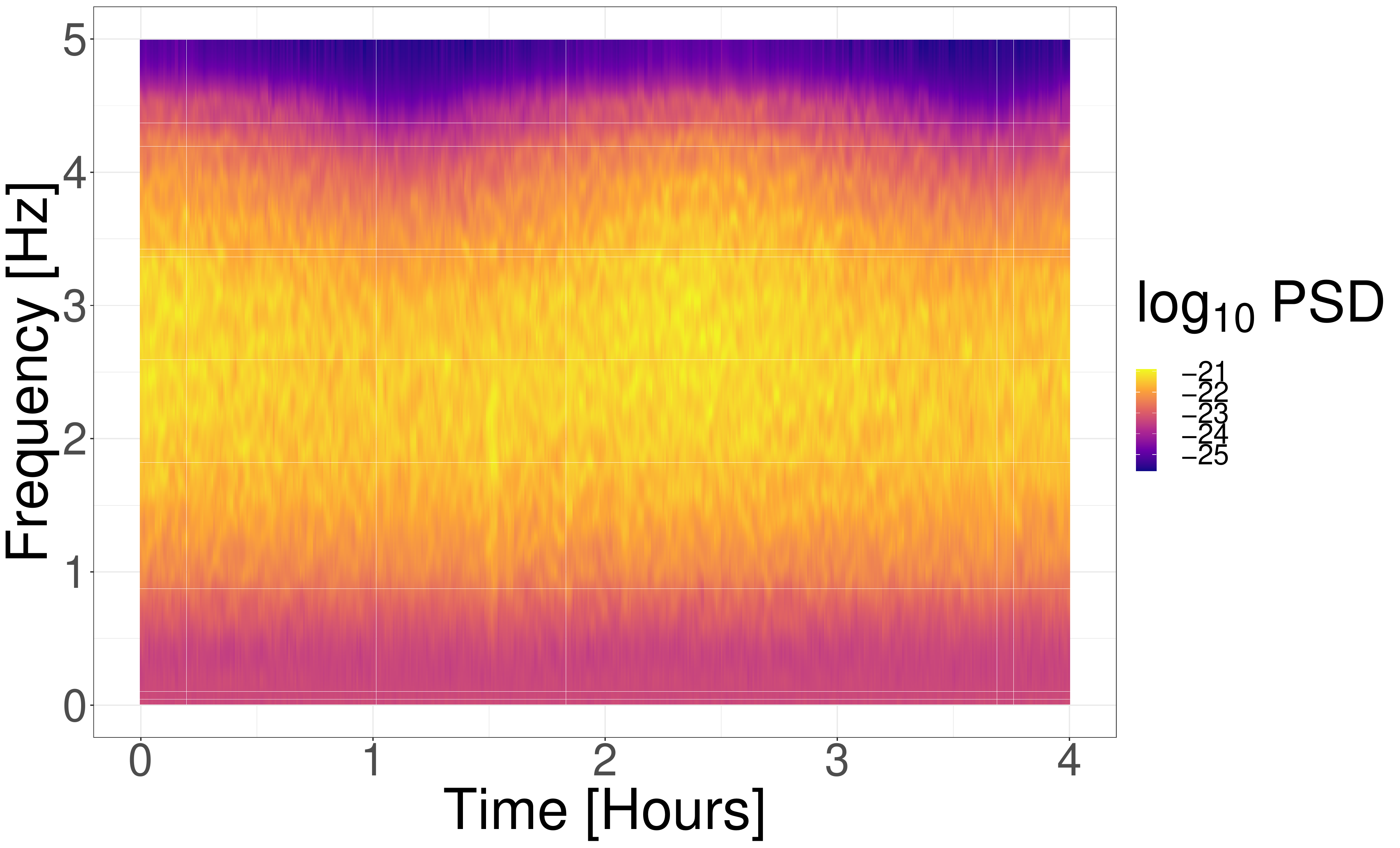}
\caption{AR Spectrogram of the AM Data Set.}
\label{fig:LPF_AM_Spectrogram}
\end{figure}

For the AM Data Set, we take the Level 3 data without any additional
preprocessing.  We examine the first four hours of this data set.
The AR spectrogram (with window length $T = 2^{10}$
  and 75\% overlap) of the AM Data Set can be seen in
  Figure~\ref{fig:LPF_AM_Spectrogram}.

\subsubsection{Glitch Data Set}

\begin{figure}[!h]
\includegraphics[width=1\linewidth]{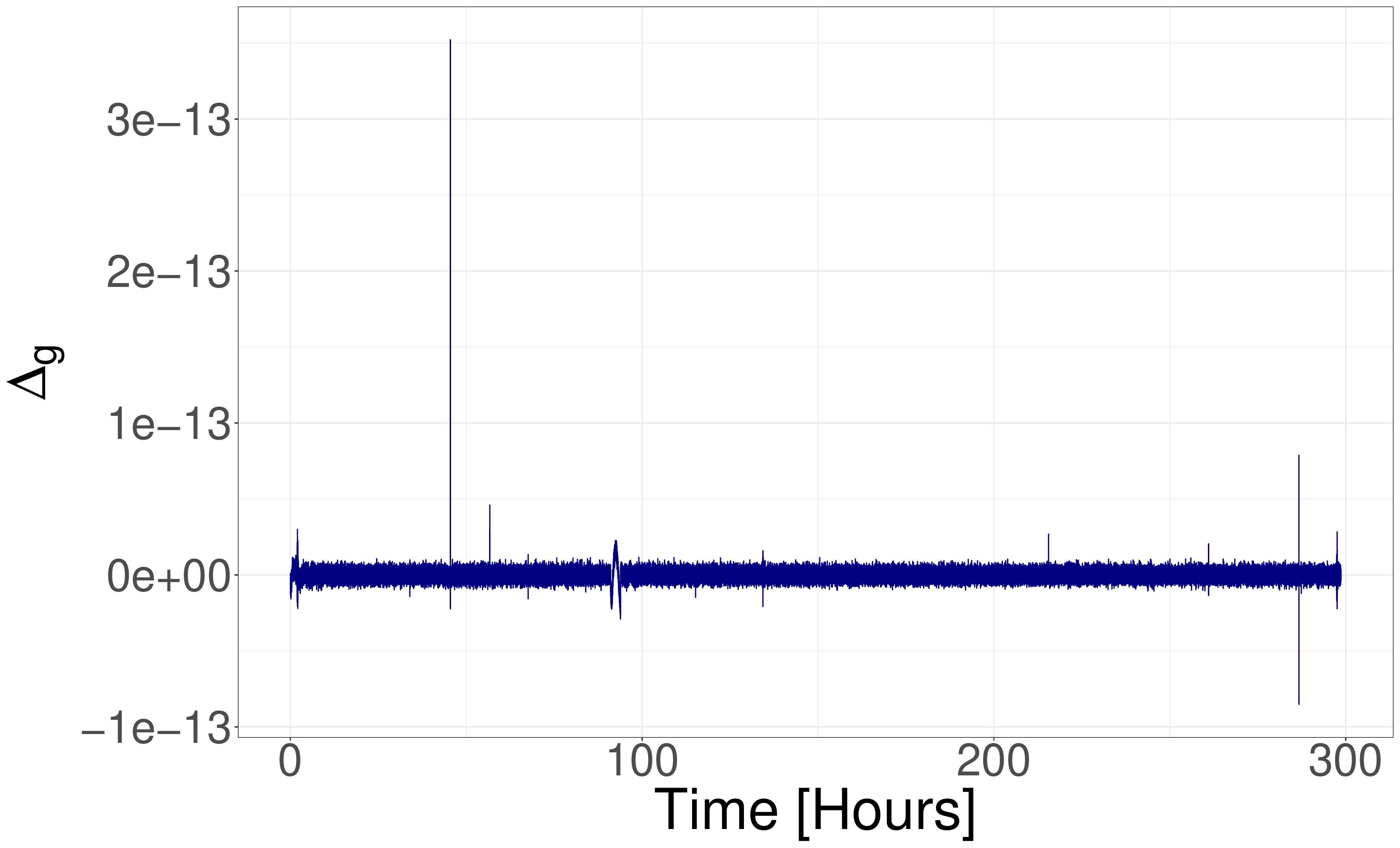}
\caption{$\Delta g$ LPF data from the Glitch Data Set.}
\label{fig:LPF_Glitch_Full}
\end{figure}

\begin{figure}[!h]
\includegraphics[width=1\linewidth]{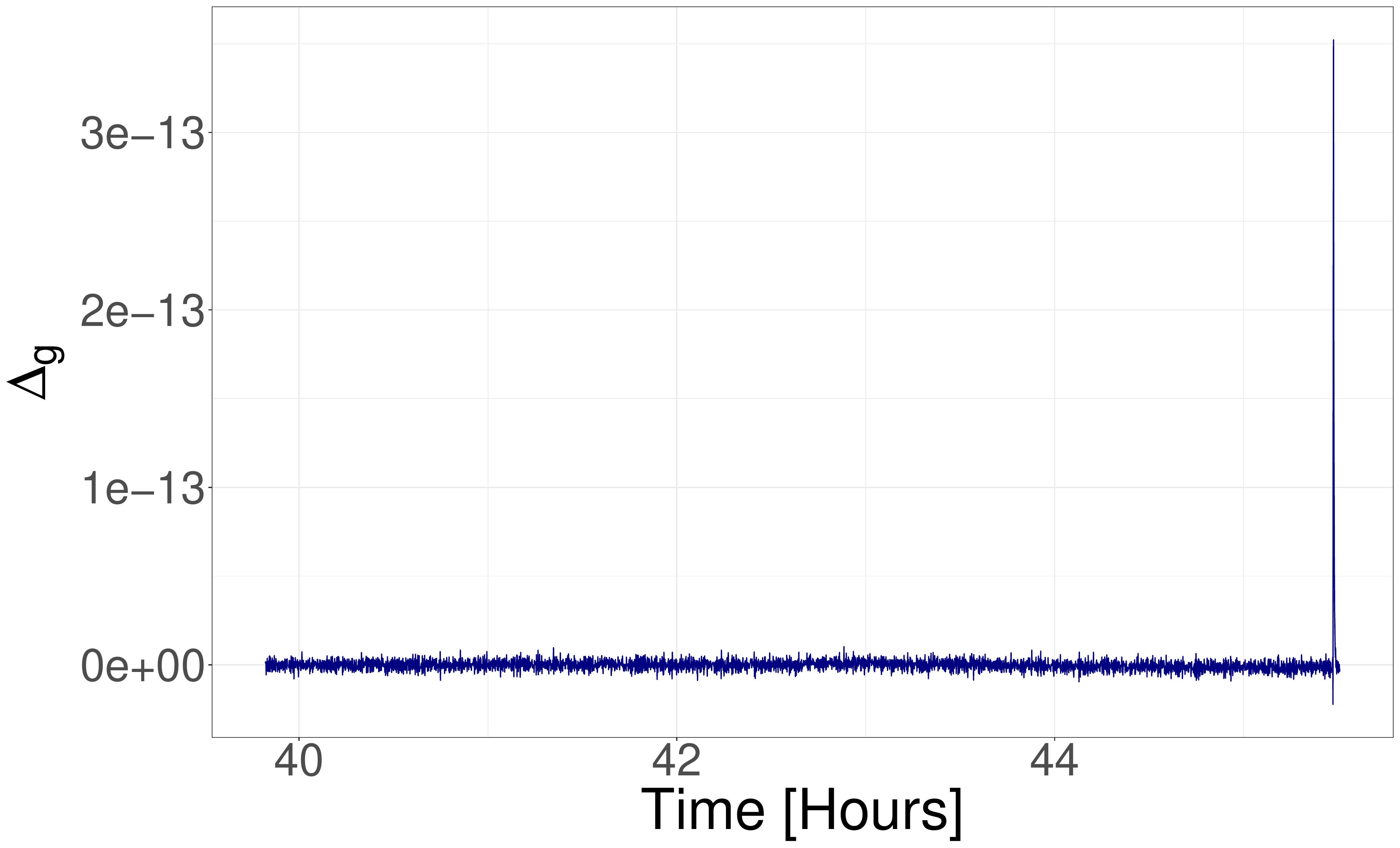}
\caption{The 14th and 15th length $n = 2^{11}$ segments from the
  Glitch Data Set.  There is a noticeably large glitch at the end of
  the displayed time series.}
\label{fig:LPF_Glitch_End}
\end{figure}

\begin{figure}[!h]
\includegraphics[width=1\linewidth]{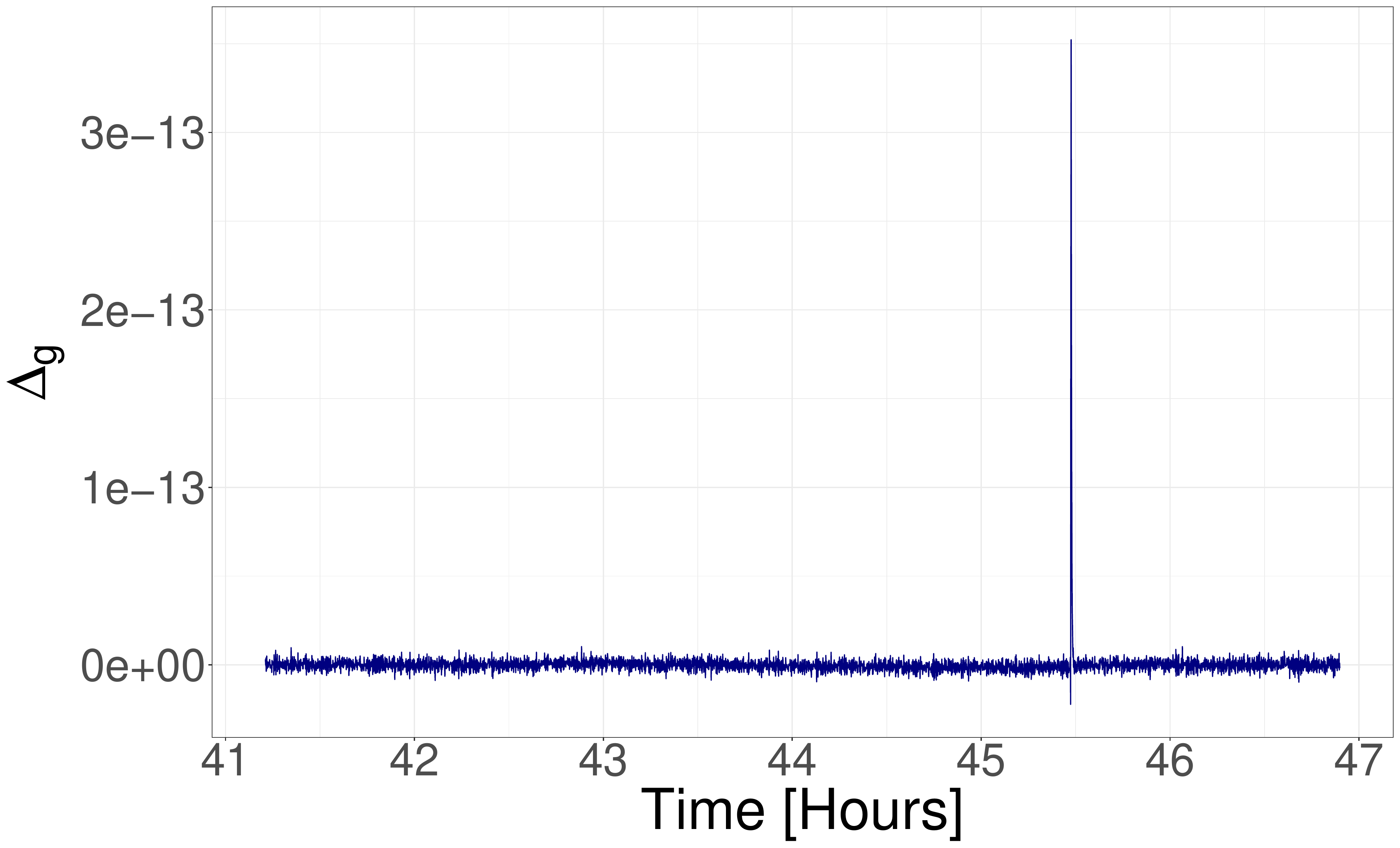}
\caption{Same data as in Figure~\ref{fig:LPF_Glitch_End} but
  translated so that the glitch occurs 75\% of the way through the
  time series.}
\label{fig:LPF_Glitch_Middle}
\end{figure}

\begin{figure}[!h]
\includegraphics[width=1\linewidth]{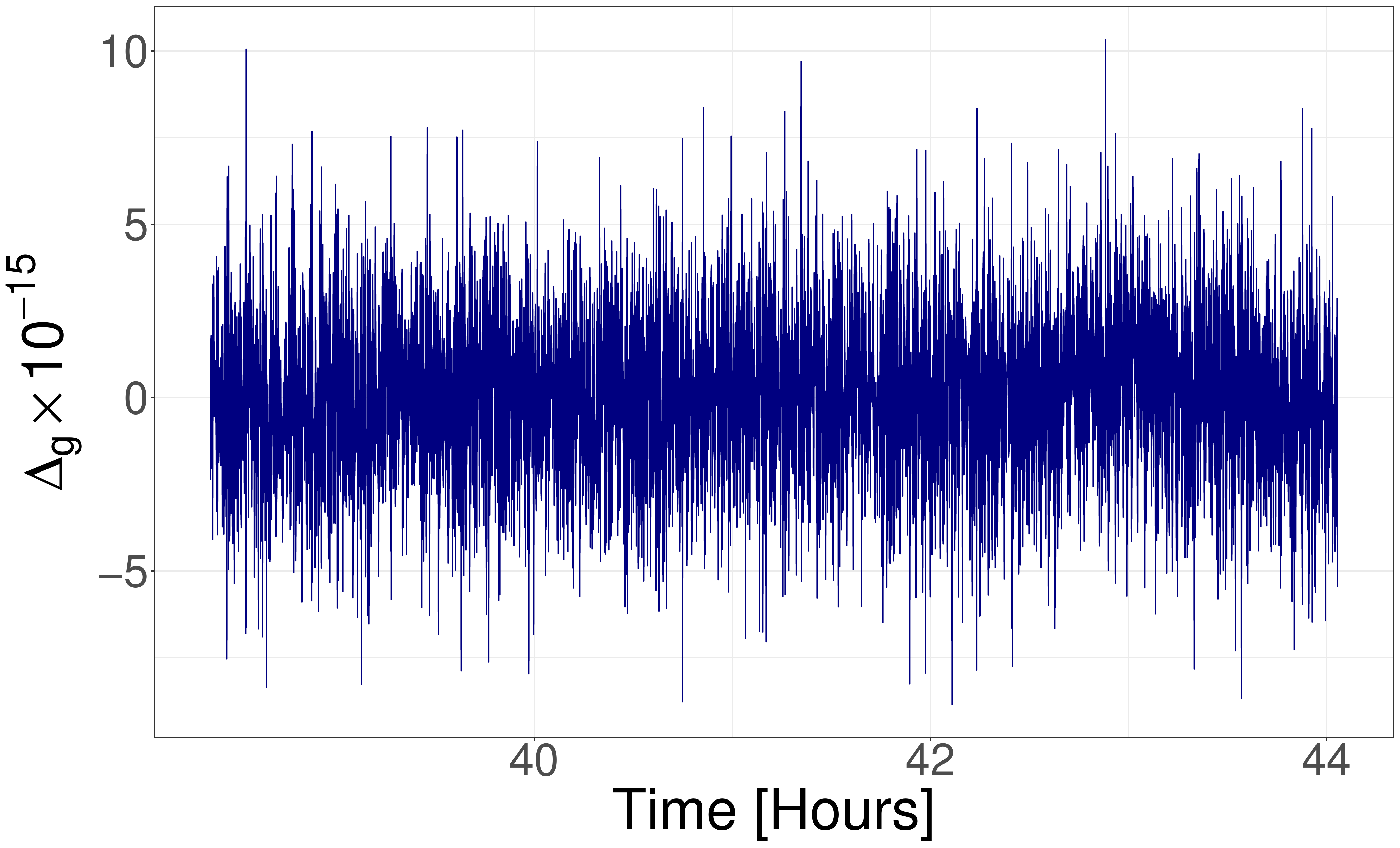}
\caption{Stationary segment of the Glitch Data Set occurring before the
  large glitch in Figures~\ref{fig:LPF_Glitch_End} and
  \ref{fig:LPF_Glitch_Middle}.}
\label{fig:LPF_Stationary}
\end{figure}

Here, we analyze the Glitch Data Set for four different cases.  These
are:
\begin{enumerate}
\item The full time series (see Figure~\ref{fig:LPF_Glitch_Full}).
\item A segment with a large glitch at the end of the time series (see
  Figure~\ref{fig:LPF_Glitch_End}).
\item A segment with a large glitch not at the end of the time series (see
  Figure~\ref{fig:LPF_Glitch_Middle}).
\item A stationary segment with no glitches present (see
  Figure~\ref{fig:LPF_Stationary}).
\end{enumerate}

For the following surrogate tests, we compute an AR spectrogram with
no overlap and window length $2^9$ for Case 1, and $2^7$ for Cases
2--4.  1,000 surrogates are then used to generate the sampling
distribution of the test statistics.


The full downsampled, filtered, and piecewise linear detrended data
can be seen in Figure~\ref{fig:LPF_Glitch_Full}.  This data set is
full of transient, high amplitude ``glitches''.

When considering the full data set, we report a $p$-value of 0.001 for
the KS variant and 0.000 for the KL and LSD variants
  of the VOCAL test, and 0.001 for the SOMED test.  These results
indicate that all of the surrogate tests provide evidence against the
notion of stationarity, which we attribute to the glitches.

Now consider the case where we look at a segment of the data set where
the largest glitch is present.  We can see in
Figure~\ref{fig:LPF_Glitch_Full} that the largest glitch in the time
series is somewhere around 45 hours into data collection (in the 15th
segment from preprocessing).  We zoom on this segment (of length $n =
2^{11}$) and its neighbouring earlier (14th) segment in
Figure~\ref{fig:LPF_Glitch_End}.

When analyzing the time series in Figure~\ref{fig:LPF_Glitch_End},
where the glitch is at the end of the time series, we report a
$p$-value of 0.001 for the KS variant of the VOCAL test,
0.000 for the KL and LSD variants of the VOCAL test,
and 0.002 for the SOMED test, all providing very strong evidence
against the notion of stationarity.  We attribute this nonstationarity
to the glitch present in the data set.

The glitch at the end of the times series causes naturally a large 
Euclidean distance for the last interval in comparison to the previous
ones in the SOMED test case.  This is reflected in the estimated simple 
regression model.  The glitch has a leverage effect in the estimated 
slope, which results in the rejection of the null hypothesis.

When the large glitch is not at the end of the time series as in
Figure~\ref{fig:LPF_Glitch_Middle}, the KS, KL, and
  LSD variants of the VOCAL test all yield $p$-values of 0.000,
meaning we have very strong evidence against stationarity.  However,
for the SOMED test, we report a $p$-value of 0.701, which means we are
not rejecting the notion of stationarity here.

Unlike the previous case, the glitch is relatively in the middle of
the sequence, which results in a large value in one of the central
cumulative medians of the Euclidean distances in the SOMED test case.
This large value has a null effect on the estimated slope of the
linear model due to its position.  Thus, the method fails wrongly to
reject the null hypothesis.  However, this large value can be
visualized via the Cook's distance, a measure of the impact of a
single observation in the parameter estimates.  In this case, the
interval that contains the glitch has a Cook's distance value of 0.39,
which is extremely close to the cut point given by the rule of thumb
0.4, and it is quite different from the rest of the Cook's distance
values, which have a median of 0.014 and standard deviation of
0.070. Even though the SOMED test fails to reject the stationary
hypothesis in this case, the glitch can be detected and thus the
validity of the conclusions based on this test can be questioned.
This procedure can be applied to other similar situations.

For Case 4 where the data looks stationary, we report the following
$p$-values: 0.836, 0.198, and 0.361 for the KS, KL,
and LSD variants of the VOCAL test respectively, and
0.702 for the SOMED test.  All three do not reject the null
hypothesis, meaning we have no evidence against stationarity for this
segment of data.

\subsubsection{AM Data Set}

\begin{figure}
\includegraphics[width=1\linewidth]{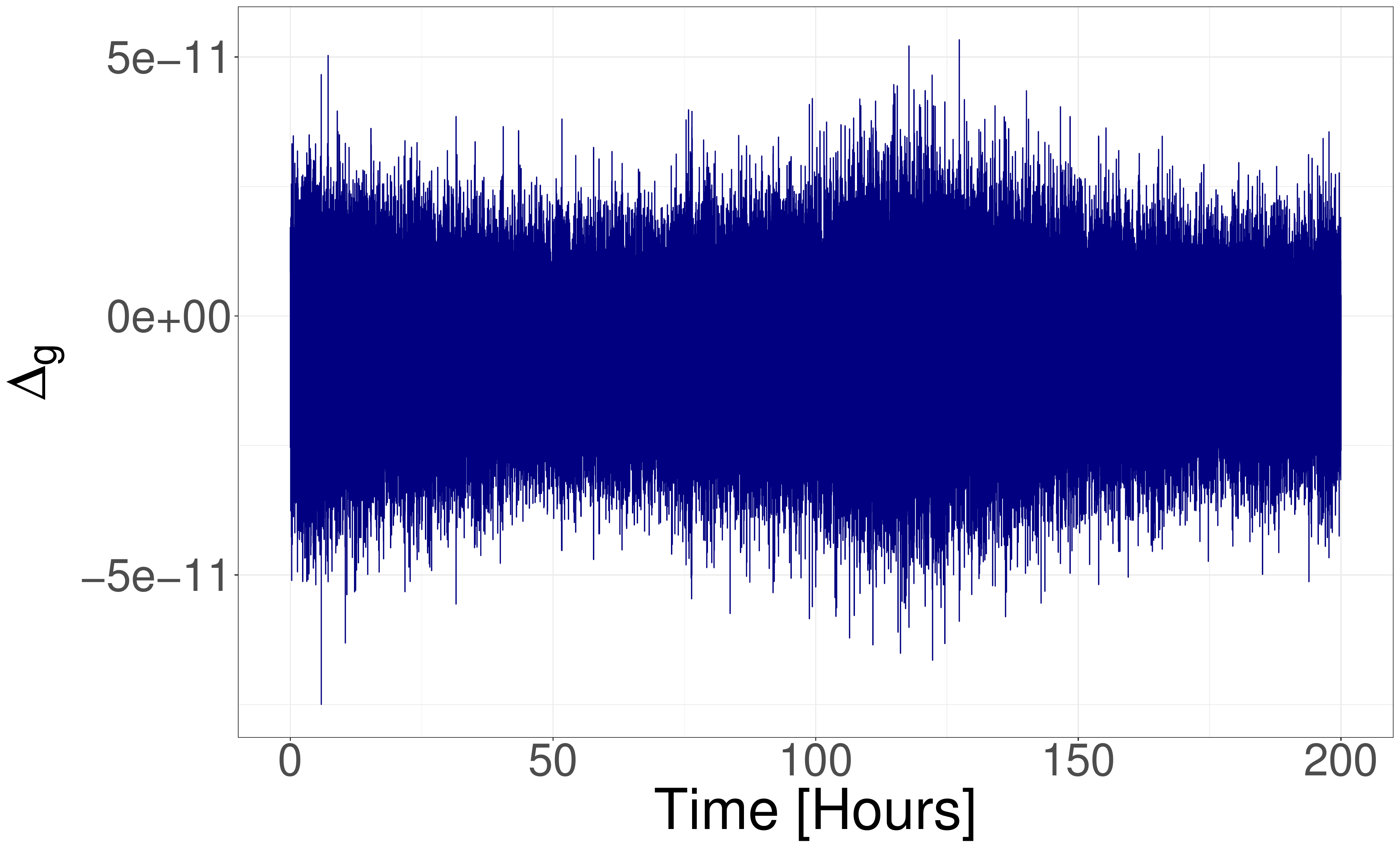}
\caption{$\Delta g$ data from the AM Data Set.}
\label{fig:LPF_Breathing}
\end{figure}

We see cyclostationary behaviour in the LPF data.  This is highlighted
in the AM Data Set, which is illustrated in
Figure~\ref{fig:LPF_Breathing}.

For all of the surrogate tests, we compute an AR spectrogram with no
overlap and window length $2^9$.  Using 1,000 surrogates to generate
the sampling distribution of the test statistics, we report a
$p$-value of 0.008 for the KS variant of the VOCAL test,
0.000 for the KL and LSD variants of the VOCAL test,
and 0.000 for the SOMED test, all providing very strong evidence
against the notion of stationarity.

\section{Addressing Nonstationary Noise}\label{sec:addressing}


Using the hypothesis tests defined in
  Section~\ref{sec:vocal} and Section~\ref{sec:somed}, or similar, we
  can identify if LISA noise is nonstationary.  This will help us to
determine where, and how often to split LISA data so that each time
segment is locally stationary, with its own noise PSD (to be
independently estimated/updated).  Once we know where to segment the
data, we can develop a LISA data analysis strategy.

Here we describe a parameter estimation routine for one non-chirping
galactic binary GW signal, where we simultaneously estimate signal
parameters and the LISA noise PSD over time to take into account the
time-varying nature of the noise.  We include a planned gap in the
data stream and use different noise structures before and after the
gap to mimic what we expect to happen to LISA noise due to
antenna repointing.


\subsection{Galactic White Dwarf Binary Gravitational Wave Signal Model}

We assume the low frequency approximation to the LISA response as
described by \citet{carre:2010}.  We define the GW strain in one
Time-Delay Interferometry (TDI) \citep{tinto:2005}
channel as
\begin{equation*}
  h(t) = h_+(t)F^+(t) + h_\times(t)F^\times(t),
\end{equation*}
where the GW polarisations are defined as
\begin{eqnarray*}
  h_+(t) &=& A_0\left(1 + \cos^2\iota\right)\cos\left(\Phi\left(t\right) + \varphi_0\right), \\
  h_\times(t) &=& -2A_0\cos\iota\sin\left(\Phi\left(t\right) + \varphi_0\right),
\end{eqnarray*}
for a non-chirping galactic white dwarf binary.  Here, $A_0$ is the
amplitude, $\iota$ is the inclination angle between the orbital plane
of the source and the observer, $\varphi_0$ is the initial phase, and
$\Phi(t)$ is the time-dependent phase, which for a circular orbit, is
defined as
\begin{equation*}
  \Phi(t) = 2\pi \omega_0\left(t + R_\oplus\sin\theta\cos\left(2\pi \omega_m t -
  \phi \right)\right),
\end{equation*}
where $\omega_0$ is the monochromatic frequency, $\omega_m$ is the LISA
modulation frequency (defined as the reciprocal of the number of
seconds in a year), $R_\oplus$ is the time light takes to travel one
astronomical unit, and $(\theta, \phi)$ is the sky location of the
source.

Using the definitions of \citet{rubbo:2004}, the antenna beam factors
are
\begin{eqnarray*}
  F^+(t) &=& \frac{1}{2}\left(\cos\left(2\psi\right)D^+\left(t\right) - \sin\left(2\psi\right)D^\times\left(t\right)\right),\\
  F^\times(t) &=& \frac{1}{2}\left(\sin\left(2\psi\right)D^+\left(t\right) + \cos\left(2\psi\right)D^\times\left(t\right)\right),
\end{eqnarray*}
where
\begin{equation*}
  \begin{split}
    D^+(t) &=
    \frac{\sqrt{3}}{64}\Bigg(-36\sin^2\left(\theta\right)\sin\left(2\alpha\left(t\right)-2\lambda\right) \\
    & + \left(3 + \cos\left(2\theta\right)\right) \bigg(\cos\left(2\phi\right)\Big(9\sin\left(2\lambda\right) \\
    & - \sin\left(4\alpha\left(t\right)-2\lambda\right)\Big) + 2\sin\left(2\phi\right) \\
    & \Big(\cos\left(4\alpha\left(t\right)-2\lambda\right) - 9\cos\left(2\lambda\right)\Big)\bigg) \\
    & -4\sqrt{3}\sin\left(2\theta\right)\Big(\sin\left(3\alpha\left(t\right)-2\lambda-\phi\right) \\
    & - 3\sin\left(\alpha\left(t\right)-2\lambda+\phi\right)\Big)\Bigg), \\
    D^\times(t) &= \frac{1}{16}\Bigg(\sqrt{3}\cos\left(\theta\right)\Big(9\cos\left(2\lambda-2\phi\right) \\
    & - \cos\big(4\alpha\left(t\right) - 2\lambda - 2\phi\big)\Big) \\
    & - 6\sin\left(\theta\right)\Big(\cos\big(3\alpha\left(t\right) - 2\lambda - \phi\big) \\
    & + 3\cos\big(\alpha\left(t\right)-2\lambda+\phi\big)\Big)\Bigg),
  \end{split}
\end{equation*}
and $\alpha(t) = 2\pi\frac{t}{T} + \kappa$ is the orbital phase of the
centre of mass of the constellation, where $T$ is the number of
seconds in a year (though in this study, we increase the orbital
modulation so that $T$ is the number of seconds in a day for
computational reasons), and $\kappa = 0$ is the initial ecliptic
longitude.

The parameters we are interested in estimating are amplitude $A_0$,
monochromatic frequency $\omega_0$, initial phase $\varphi_0$, and
inclination $\iota$. All other parameters, e.g., sky location
($\theta, \phi$), GW polarization angle $\psi$, and initial ecliptic
longitude $\kappa$, are fixed.  To this end, we place the following
noninformative priors on the signal parameters:
\begin{eqnarray*}
  A_0 &\sim& \mathrm{Uniform}[0, \infty), \\
    \cos\varphi_0 &\sim& \mathrm{Uniform}[-1, 1], \\
    \cos\iota &\sim& \mathrm{Uniform}[-1, 1], \\
    \omega_0 &\sim& \mathrm{Uniform}[0.0001, 0.0191].
\end{eqnarray*}

Although data will eventually be analyzed in the three TDI channels A,
E, and T \citep{tinto:2005} (where T is the noise-only channel
containing no signal information), for simplicity, we will only
consider the A channel, meaning we set TDI channel angle $\lambda =
0$.

\subsection{Bayesian Nonparametric Noise Model}\label{sec:bspline}

To model the noise PSD, we use the Bayesian nonparametric B-spline
prior introduced by \citet{edwards:2017}.  The B-spline prior has the
following representation as a mixture of B-spline densities:
\begin{equation*}
  s_r(x; k, \w_k,\boldsymbol\xi) = \sum_{j = 1}^k w_{j,k}
  b_{j,r}(x; \boldsymbol\xi),
\end{equation*}
where $b_{j,r}(.)$ is the $j^{\mathrm{th}}$ B-spline density of fixed
degree $r$, $k$ is the number of B-spline densities in the mixture,
$\w_k=(w_{1,k},\ldots,w_{k,k})$ is the weight vector, and
$\boldsymbol\xi$ is the nondecreasing knot sequence.

The noise PSD $f(.)$ is then modelled as as follows:
\begin{equation*}
  f(\pi x) = \tau \times s_r(x; k, G, H), \quad x \in [0, 1],
\end{equation*}
where the mixture weights and knot differences are induced by CDFs $G$
and $H$ respectively, each on $[0, 1]$, and $\tau = \int_0^1
f(\pi x) \mathrm{d} x$ is the normalization constant.

We place the following \textit{a priori} independent priors on the
noise PSD model parameters $(k, G, H, \tau)$:
\begin{eqnarray*}
  p(k) &\propto& \exp\{-\theta k^2\},\\
  G &\sim& \mathrm{DP}(G_0, M_G), \\
  H &\sim& \mathrm{DP}(H_0, M_H), \\
  \tau &\sim& \mathrm{IG}(\alpha, \beta),
\end{eqnarray*}
where DP represents a Dirichlet process, IG is the inverse-gamma
distribution, $\theta$ is a smoothing coefficient, $G_0$ and $H_0$ are
base measures, and $M_G$ and $M_H$ are concentration parameters.

Finally, the joint prior is updated by the commonly used Whittle
likelihood \citep{whittle:1957} to yield a pseudo-posterior.  For more
details, such as implementation, we refer the reader to
\citet{edwards:2017}.

This is in essence a blocked Metropolis-within-Gibbs sampler similar
to \citet{edwards:2015}, where we iteratively sample the signal
parameters given the noise parameters, and then the noise parameters
given the signal parameters and so on.


Ignoring galactic confusion noise, the LISA
  sensitivity curve in the A TDI channel as defined by
  \citet{ldc_manual, karnesis:2020} is:
  \[
  \begin{split}
    S_A(x) &= 8\sin^2(x) \times \biggl(P_{\text{OMS}} \times \bigl(2 +
    \cos(x)\bigr) \\ &\quad + 2 \times P_{\text{Acc}} \times \bigl(3 +
    2\cos^2(x) + \cos(2x)\bigr)\biggr)
  \end{split}
  \]
 where $x = 2\pi f L / c$, $f$ is frequency in Hertz, $c$ is the speed
 of light, $L$ is the satellite arm length ($2.5 \times 10^9$ metres).
 $P_{\text{OMS}}$ is optical metrology noise, defined as:
  \begin{equation*}
    P_{\text{OMS}} = (1.5 \times 10^{-11})^2 \left(1 +
    \left(\frac{2 \times 10^{-3}}{f}\right)^4\right)\left(\frac{2\pi
      f}{c}\right)^2.
  \end{equation*}
  Acceleration noise $P_{\text{Acc}}$ is defined as follows:
  \[
  \begin{split}
    P_{\text{Acc}} &= (3 \times 10^{-15})^2 \biggl(1 +
    \biggl(\frac{4\times10^{-4}}{f}\biggr)^2 \biggr) \times \\
    &\quad \biggl(1 + \biggl(\frac{f}{8\times10^{-3}}\biggr)^4\biggr)
    \left(2\pi f c \right)^{-2}.
  \end{split}
  \]
  These terms are constructed in \citet{robson:2019}.  We can then
  easily simulate Gaussian noise, coloured by $S_A(.)$.

\subsection{Example}

Consider the simple case where we have 48 hours of data from the A TDI
LISA channel, and there is one planned outage at 22 hours for a
duration of four hours due to antenna repointing.  Assume this antenna
repointing changes the noise structure.  Whether this is realistic is
yet to be determined.

We generate a (non-chirping) galactic white-dwarf binary signal with
the following parameters to be estimated:
\begin{eqnarray*}
  A_0 &=& 1 \times 10^{-21} \\
  \omega_0 &=& 0.005 \\
  \varphi_0 &=& 3\pi / 4 \\
  \iota &=& \pi / 2.
\end{eqnarray*}
We fix the sky location $(\theta = \pi / 4, \psi = \pi / 4)$ and GW
polarization angle $\phi = 0$.  Let TDI channel angle $\lambda =
0$ as we only consider the A channel.  We set the sample interval to
$\Delta_t = 10~\mathrm{s}$, yielding a Nyquist frequency of $\omega_{*} =
0.05~\mathrm{Hz}$.

The noise for this example is created as follows.
Before the gap, we generate Gaussian noise, coloured
  by the LISA sensitivity curve in the A TDI channel, $S_A(.)$.  After
  the gap, we generate Gaussian noise, coloured by an ``optical
  metrology noise modified'' version of the LISA sensitivity curve in
  the A channel.  We adjust the scale and shape of the of the optical
  component of the noise.  Instead of using $P_{\text{OMS}} \times
  \cos(2 + x)$, we use $2P_{\text{OMS}} \times \cos(2 + 2x)$, thus
  adjusting the scale and shape of the optical metrology component.
The increase in the variance of noise and the change in the
autocovariance structure during the second half is our attempt at
simulating a change in noise structure due to the repointing of
antennae.  This noise setup yields an overall signal-to-noise ratio
(SNR) of $\varrho \approx 50$ (when considering both noise segments).


We add this noise to the generated GW signal and remove the middle
four hours of the data to create a gap.  We then multiply the data by
a Tukey-type window, where we taper off any data to zero where there
is a gap, with a chosen taper parameter of $r = 0.01$.  Note that this
Tukey-type window will be applied to all galactic white-dwarf binary
signals proposed during the MCMC algorithm to ensure gaps are in the
correct place in the signal model.

\begin{figure}[h]
\includegraphics[width=0.9\linewidth]{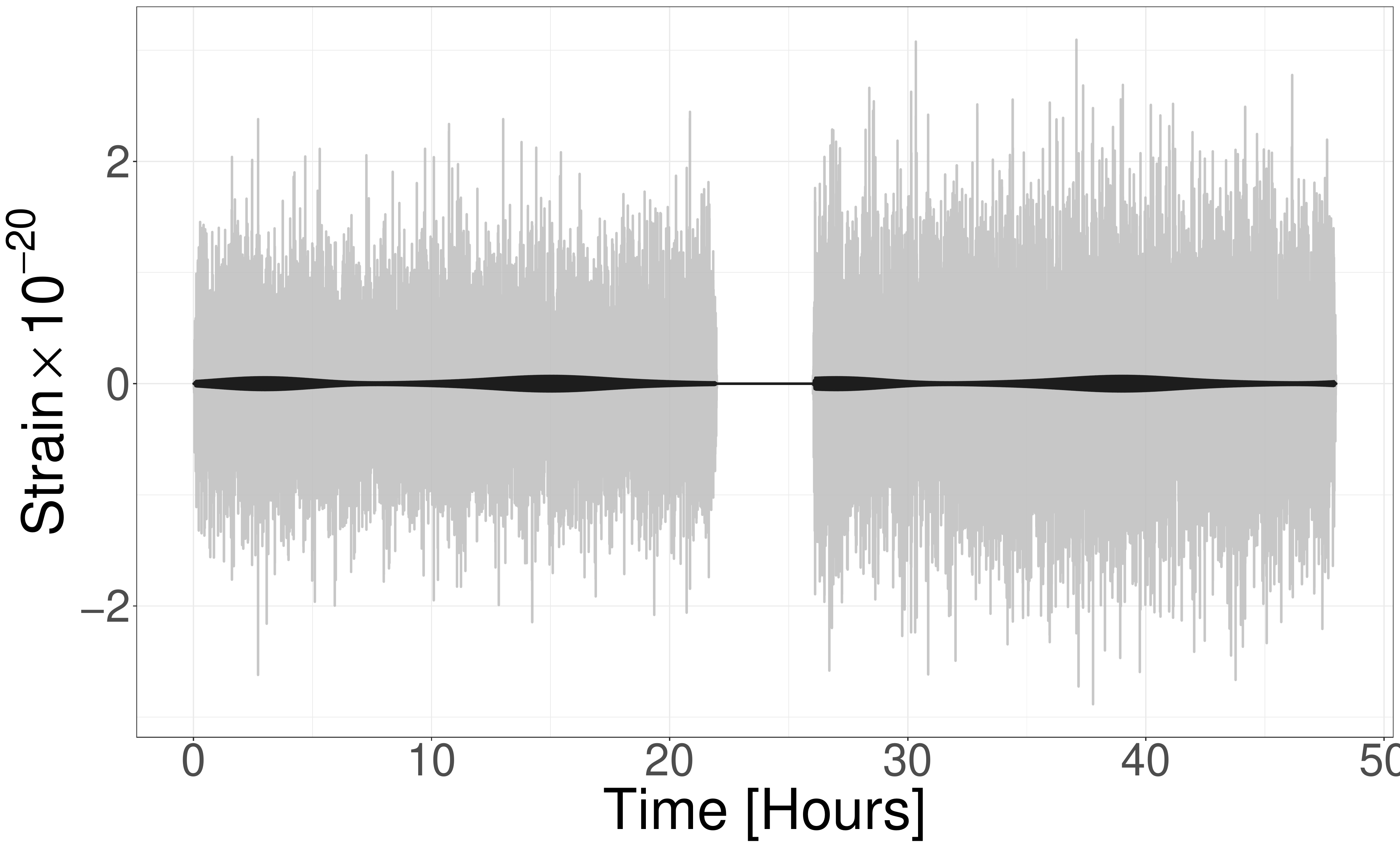}
\caption{Non-chirping galactic white-dwarf binary GW
    signal (black) and signal plus noise (grey).  A four hour gap is
    inserted in the middle, multiplied by Tukey-type window (with $r =
    0.01$).  The first half of the noise series is generated using the
    LISA sensitivity curve in the A TDI channel and the second half is
    generated using an optical metrology noise modified version of
    this.}
\label{fig:wdwd_gap_setup}
\end{figure}

A realization of this data setup can be seen in
Figure~\ref{fig:wdwd_gap_setup}. 

We conduct parameter estimation with the assumption
  of piecewise stationary noise.  This allows us to model the noise PSD before
and after the gap differently if they are in fact different (which
they are in this example).  Even if the noise was
  stationary, there would be no harm conducting analysis this way.  A
  model that allows for a time-varying noise PSD mitigates against
  possible parameter estimation biases caused by assuming noise is
  stationary.  We model the two noise PSDs  using two independent
 nonparametric  B-spline priors presented in Section~\ref{sec:bspline}.


\subsection{Results and Model Checking}

We run the MCMC algorithm for 100,000 iterations, with a burn-in of
50,000 and thinning factor of 5.  We also use an adaptive proposal for
each signal parameter described by \citet{roberts:2009}.  That is, for
each parameter, we use a standard Metropolis step with Normal proposal
centred on the previous value, and variance that is automatically
tuned to achieve a desired acceptance rate of 0.44.

\begin{figure}[h]
\includegraphics[width=1\linewidth]{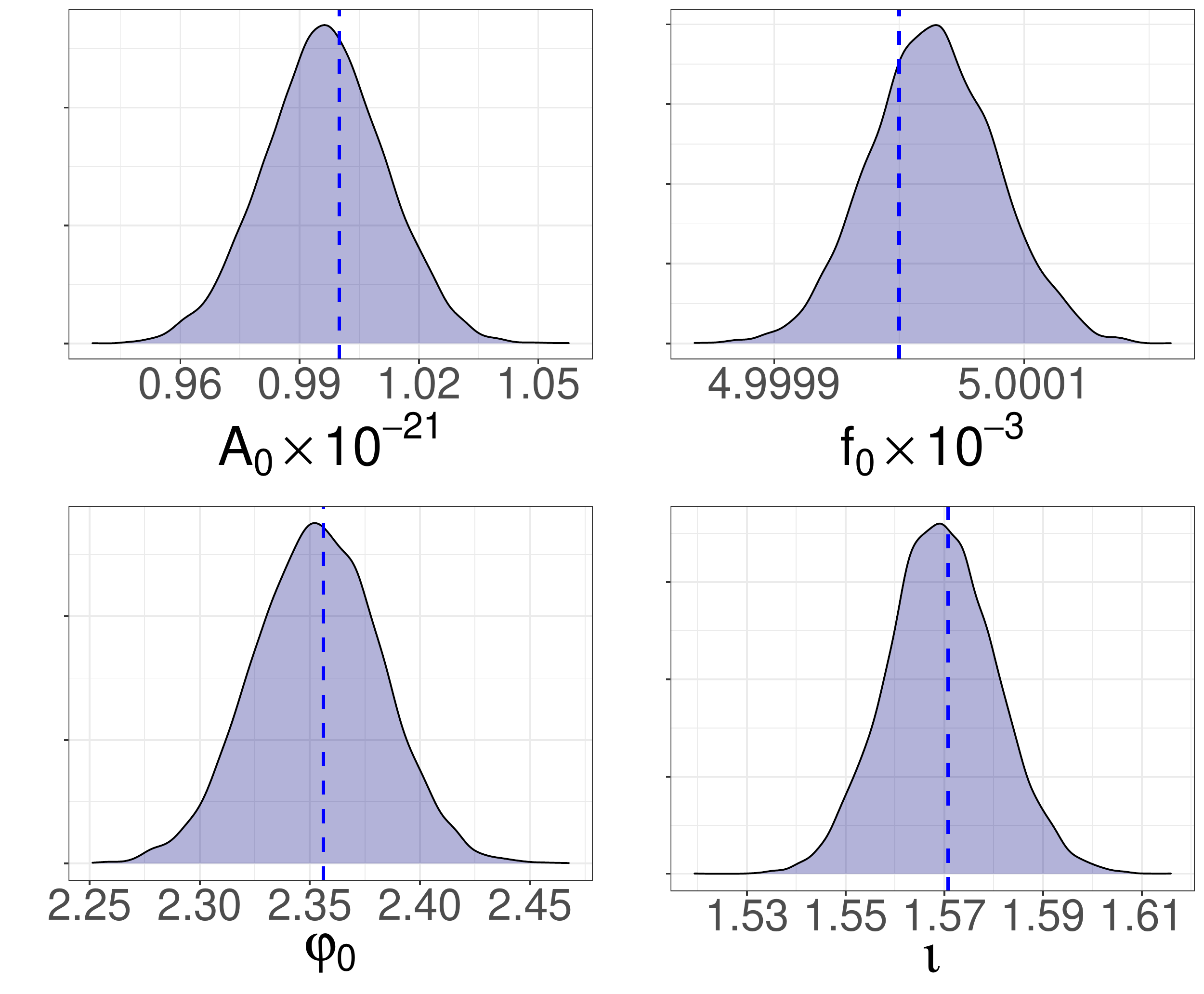}
\caption{Posterior densities for the galactic white dwarf binary
  parameters.  The dashed vertical line is the true parameter.}
\label{fig:Posteriors}
\end{figure}

As illustrated in Figure~\ref{fig:Posteriors}, we can
  accurately recover the GW signal parameters in the presence of
  nonstationary noise due to a simulated planned gap that changes the
  optical contribution to LISA noise.

\begin{figure}[h]
\includegraphics[width=1\linewidth]{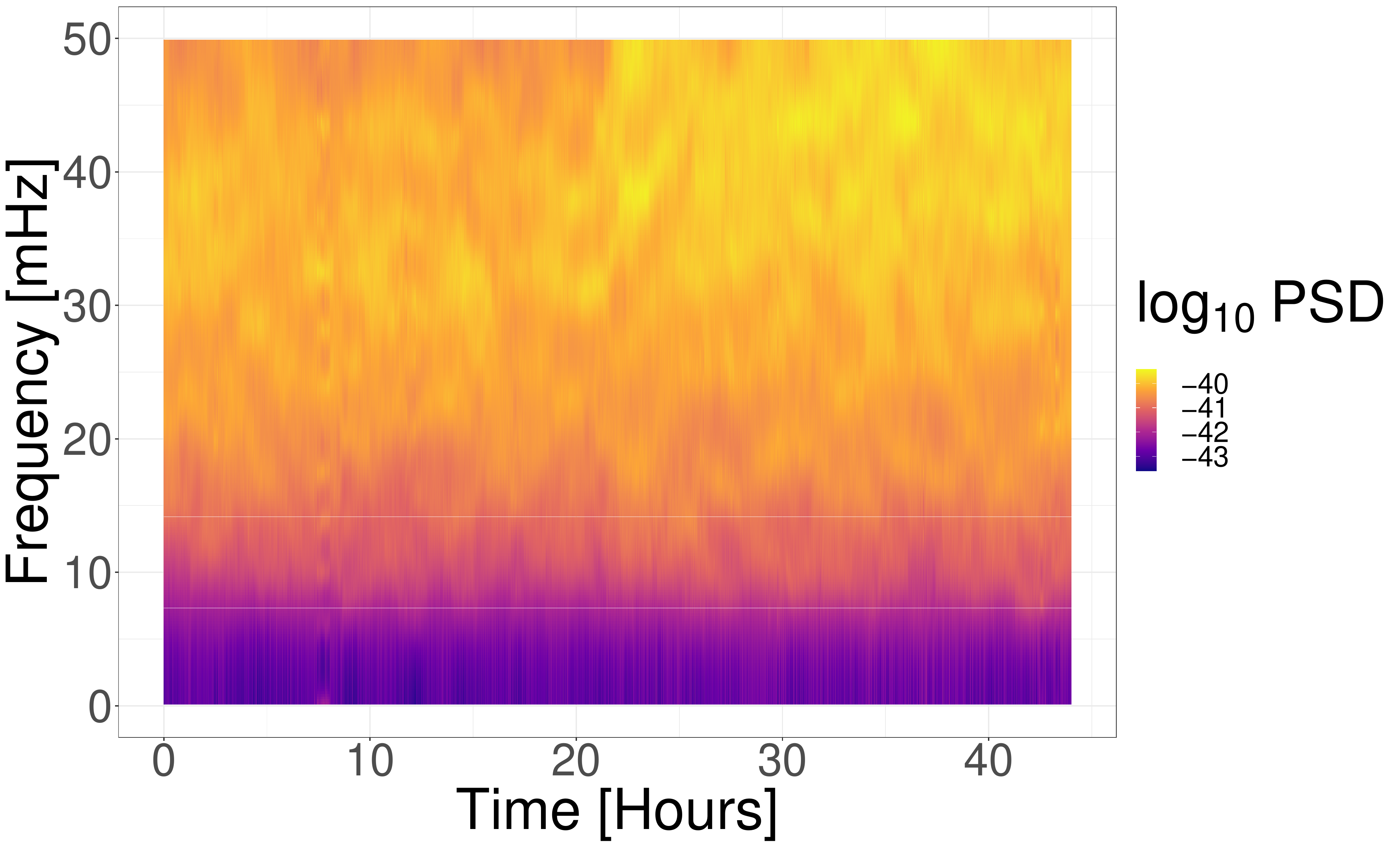}
\caption{AR spectrogram of residuals after removing
    the posterior median signal from the data.  There is a noticeable
    change in power at the high frequencies in the second half of the
    spectrogram.}
\label{fig:Residuals}
\end{figure}

{\em Model checking}, i.e.\ a careful investigation of
  the correctness of any model assumptions, should be part of all
  statistical inference procedures. To check whether it was
  appropriate to assume that the individual time series before and
  after the gap were in fact stationary, we can apply the stationarity
  tests based on the surrogate data approach to the time series of
  residuals before and after the gap. Moreover, to check whether we
  could have safely assumed that the full time series is stationary
  and thus potentially enabled an analysis with one single B-spline
  prior for the noise component instead of two different noise models,
  we apply the stationarity test to the residuals of the full time
  series.  The residual time series can be thought of as the ``best
  guess'' of underlying noise.  We calculate the posterior median GW
  signal and subtract this from the data to compute the residual
  series, and then concatenate the residuals before and after the gap.
  The AR spectrogram of these residuals is highlighted in
  Figure~\ref{fig:Residuals}.  Running the surrogate tests on the
  residuals, we report $p$-values (assuming a window length of $T =
  2^9$ and overlap of 75\%) in Table~\ref{tab:pvalues}.  For all
  variants of the surrogate test, we may reject the notion of
  stationarity for the full residual time series.  We also do not
  reject the hypothesis of stationarity for the first and second
  halves. This confirms that our stationarity assumptions for each
  time series before and after the gap was justified and that it was
  appropriate to assume two different nonparametric noise models.

\begin{table}[!h]
    \begin{center}
      \caption{\label{tab:pvalues} $p$-values of the
          surrogate tests for the residual time series using a window
          length of $T= 2^9$ and overlap 75\%.}
    \begin{tabular}{c|cccc}
    \hline
    &KS&KL&LSD&SOMED\\
    \hline
    Full Series&0.000&0.000&0.001&0.000 \\
    Before Gap&0.189&0.492&0.597&0.580 \\
    After Gap&0.488&0.445&0.934& 0.367\\
    \hline
    \end{tabular}
  \end{center}
\end{table}

\section{Discussion}\label{sec:discussion}

In this paper, we have discussed methods to identify and address
nonstationary noise in the future LISA mission.  We demonstrated the
usefulness of the lesser-known nonparametric surrogate tests for assessing the
stationarity of a time series, introducing a novel variant in the form
of the SOMED test.  We applied the surrogate tests to real LPF data
and showed that certain segments are nonstationary in nature, due to
glitches, and amplitude modulations.  As the architecture of LISA will
share many similarities to LPF, we see this as an important first step
in understanding the stationarity/nonstationarity of LISA data.

We introduced a Bayesian semiparametric framework for conducting
parameter estimation when there is nonstationary noise as a result of
antenna repointing.  Assuming a stationary
noise model in this situation may lead to systematic biases in
astrophysical parameter estimates, as well as larger posterior
variances as has been investigated by \cite{ChatziioannouKaterina2019Nsem,TalbotColm2020Gawa,BiscoveanuSylvia2020QtEo}.

An interesting alternative framework for modelling piecewise stationary noise could be to
modify the time-varying spectrum estimation regime of
\citet{rosen:2012}, which utilizes reversible jump MCMC
\citep{green:1995} to determine the number of locally stationary
segments in a time series.  One could use a blocked
Metropolis-within-Gibbs sampler similar to the one introduced in this
paper to model signal parameters given noise parameters and vice
versa.  This is one avenue we aim to explore in a future paper.

Another future initiative includes investigating the impact of planned
data gaps and nonstationary noise on EMRI GW signals, particularly
those arising from near-extremal black holes.

\section*{Acknowledgements}

We thank the New Zealand eScience Infrastructure (NeSI) for their high
performance computing facilities, and the Centre for eResearch at the
University of Auckland for their technical support.  ME's and JG's
work is supported by UK Space Agency grant ST/R001901/1.  PM's and
RM's work is supported by Grant 3714568 from the University of
Auckland Faculty Research Development Fund and the DFG Grant KI
1443/3-1. RM gratefully acknowledges support by a James Cook
Fellowship from Government funding, administered by the Royal Society
Te Ap\={a}rangi.  NC and NK were supported by the Centre national
d'\'{e}tudes spatiales (CNES).  We also thank the LISA Pathfinder
Collaboration for providing us with the data sets used in this
manuscript.

\bibliographystyle{spbasic}
\bibliography{refs}  

\end{document}